\documentclass{aa}  

\usepackage{graphicx}
\usepackage{lipsum}

\usepackage[colorlinks]{hyperref}
\usepackage{color, colortbl} %ADDED BY JB
\usepackage{longtable} %ADDED BY JB
\usepackage{lscape}
\usepackage{txfonts}
\newcommand{\cii}{[C\,{\sc ii}] }
\newcommand{\ciii}{C\,{\sc iii}] }
\newcommand{\oiii}{[O\,{\sc iii}] }

\newcommand{\kms}{\,km\,s$^{-1}$}
\newcommand{\lcii}{$L_{\rm [CII]}$}
\newcommand{\lya}{Ly-$\alpha$}

\begin{document}

   \title{ALMA Lensing Cluster Survey: A spectral stacking analysis of \cii in lensed $z\sim6$ galaxies}

   %\subtitle{bla}

   \author{Jean-Baptiste Jolly\inst{1} \and Kirsten Knudsen\inst{1} \and Nicolas Laporte\inst{2,3} \and Johan Richard\inst{4} \and Seiji Fujimoto\inst{5,6} \and Kotaro Kohno\inst{7,8} \and Yiping Ao\inst{9,10} \and Franz E. Bauer\inst{11,12} \and Eiichi Egami\inst{13} \and Daniel Espada\inst{14,15} \and Miroslava Dessauges-Zavadsky\inst{16} \and Georgios Magdis\inst{5,6,17} \and Daniel Schaerer\inst{16,18} \and Fengwu Sun\inst{13} \and Francesco Valentino\inst{5,6} \and Wei-Hao Wang\inst{19} \and Adi Zitrin\inst{20}}

   \institute{Department of Space, Earth and Environment, Chalmers University of Technology, Onsala Space Observatory, SE-439 92 Onsala, Sweden\\
              \email{jean.jolly@chalmers.se}
         \and
        Kavli Institute for Cosmology, University of Cambridge, Madingley Road, Cambridge CB3 0HA, UK
        \and
        Cavendish Laboratory, University of Cambridge, 19 JJ Thomson Avenue, Cambridge CB3 0HE, UK 
        \and
        CRAL, Observatoire de Lyon, Universit\'e Lyon 1, 9 Avenue Ch. Andr\'e, F-69561 Saint Genis Laval Cedex, France
        \and 
        Cosmic Dawn Center (DAWN), Denmark 
        \and
        Niels Bohr Institute, University of Copenhagen, Jagtvej 128 DK-2200 Copenhagen N, Denmark
        \and
        Institute of Astronomy, Graduate School of Science, The University of Tokyo, 2-21-1 Osawa, Mitaka, Tokyo 181-0015, Japan
        \and
        Research Center for the Early Universe, Graduate School of Science, The University of Tokyo, 7-3-1 Hongo, Bunkyo-ku, Tokyo 113-0033, Japan
        \and
        Purple Mountain Observatory \& Key Laboratory for Radio Astronomy, Chinese Academy of Sciences, Nanjing, China
        \and
        School of Astronomy and Space Science, University of Science and Technology of China, Hefei, Anhui, China
        \and
        Instituto de Astrofisica, Facultad de Fisica, Pontificia Universidad Catolica de Chile Av. Vicuna Mackenna 4860, 782-0436 Macul,Santiago, Chile
        \and
        Millennium Institute of Astrophysics (MAS), Nuncio Monse nor Santero Sanz 100, Providencia, Santiago, Chile
        \and
        Steward Observatory, University of Arizona, 933 N. Cherry Avenue, Tucson, 85721, USA
        \and
        SKA Organisation, Lower Withington, Macclesfield, Cheshire SK11 9DL, UK
        \and
        Departamento de F\'{i}sica Te\'{o}rica y del Cosmos, Campus de Fuentenueva, Universidad de Granada, E18071-Granada, Spain
        \and 
        Observatoire de Gen\`{e}ve, Universit\'{e} de Gen\`{e}ve, 51 Ch. des Maillettes, 1290 Versoix, Switzerland
        \and
        DTU-Space, Technical University of Denmark, Elektrovej 327, DK-2800 Kgs. Lyngby, Denmark
        \and 
        CNRS, IRAP, 14 Avenue E. Belin, 31400 Toulouse, France
        \and
        Institute of Astronomy and Astrophysics, Academia Sinica, Taipei 10617, Taiwan
        \and
        Physics Department, Ben-Gurion University of the Negev, P.O. Box 653, Be'er-Sheva 84105, Israel }

   \date{Received MMMM DD, YYYY; accepted MMMM DD, YYYY}

% \abstract{}{}{}{}{} 
% 5 {} token are mandatory
 
  \abstract
  % context heading (optional)
  % {} leave it empty if necessary  
  {The properties of galaxies at redshift $z>6$ hold the key to our understanding of the early stages of galaxy evolution and can potentially identify the sources of the ultraviolet radiation that give rise to the epoch of reionisation. 
  The far-infrared cooling line of \cii at 158\,$\mu$m is known to be bright and correlate with the star formation rate (SFR) of low-redshift galaxies, and hence is also suggested to be an important tracer of star formation and interstellar medium properties for very high-redshift galaxies. 
  }
  % aims heading (mandatory)
   {
   With the aim to study the interstellar medium properties of gravitationally lensed galaxies at $z>6$, we search for \cii and thermal dust emission in a sample of 52 $z\sim6$ galaxies observed by the ALMA Lensing Cluster Survey (ALCS).
   }
  % methods heading (mandatory)
   {
   We perform our analysis using \textsc{LineStacker}, stacking both \cii and continuum emission. The target sample is selected from multiple catalogues, and the sample galaxies have spectroscopic redshift or low-uncertainty photometric redshifts ($\sigma_z < 0.02$) in nine galaxy clusters. Source properties of the target galaxies are either extracted from the literature or computed using spectral energy distribution (SED) fitting. Both weighted-average and median stacking are used, on both the full sample and three sub-samples.
   }
  % results heading (mandatory)
   {Our analyses find no detection of either \cii or continuum. An upper limit on $L_{\rm [CII]}$ is derived, implying that \cii remains marginally consistent for low-SFR $z>6$ galaxies but likely is under-luminous compared to the local \lcii-SFR relationship. 
   We discuss potential biases and possible physical effects that may be the cause of the non-detection. Further, the upper limit on the dust continuum implies that less than half of the star formation is obscured. }
  % conclusions heading (optional), leave it empty if necessary 
   {}


   \maketitle
%
%-------------------------------------------------------------------

\section{Introduction}

The first billion years after the Big Bang mark the epoch of early galaxy formation and evolution. The ultraviolet (UV) photons from the early galaxies ionised the hydrogen of the intergalactic medium, a period also known as the epoch of reionisation (EoR; see \citealt{Stark2016} for a complete review). 

Dedicated efforts to search for $z>6$ galaxies employing a variety of multi-wavelength data have resulted in thousands of viable candidates. The primary methods for discovering these early galaxies are photometric drop-out techniques and narrow-band Lyman-alpha emitter selection \citep[e.g.][]{Steidel1996,Ono2012,Ono2018,Schenker2012,Oesch2015,Zitrin2015b,Song2016,Shibuya2018,Higuchi2019}. The evolution of the derived luminosity function as a function of redshift for rest-frame UV-selected galaxies shows an increasing contribution to the total ionising UV light produced by sub-$L^*$ galaxies \citep{Bouwens2006,Bouwens2015,Mclure2013}. This has implications for understanding which sources might be responsible for the largest fraction of the reionisation of the intergalactic medium during the EoR \citep{Bouwens2007,Bouwens2012,Ouchi2009,Bunker2010,McLure2010,Oesch2010,Robertson2013,Atek2015}.

Spectroscpic redshift surveys have primarily targeted galaxies brighter than $L^*$ \citep[e.g.][]{Lilly2009,Kochnek2012,Newman2013,Stark2013}; however, the number of systematic spectroscopic redshift measurements for fainter galaxies is growing \citep[e.g.][]{LeFevre2013,Bacon2017,Hasinger2018,Coe2019,Richard2020}. Spectroscopic redshifts are essential for detailed investigations of the interstellar medium (ISM) and thus for understanding the physical conditions of the gas and dust. In the local universe the far-infrared fine-structure line of \cii at 158\,$\mu$m from the $^2P_{3/2} \to ^2P_{1/2}$ transition is known to be one of the brightest gas-cooling lines in star forming galaxies \citep{DeLooze2014,Sargsyan2014,Cormier2015}. As carbon has an ionisation potential of 11.2\,eV, below that of hydrogen, \cii is detected in regions of both predominantly neutral and ionised gas. The line luminosity has been found to correlate with the star formation rate (SFR) in normal, star-forming galaxies \citep{Stacey2010,DeLooze2011,DeLooze2014,Herrera-Camus2015,Schaerer2020} and has also been proposed as a gas mass tracer \citep[e.g.][]{Zanella2018,madden20}. As the emission wavelength is redshifted into the submillimetre and millimetre bands, the \cii is recognised as an important probe of the ISM of high-$z$ galaxies.  With an excitation temperature of $T = 91.2$\,K, the line is less sensitive to the effects of the cosmic microwave background radiation compared to, for example, CO lines, which are otherwise common tracers of molecular gas and the ISM of high-$z$ galaxies. Notably, several studies have observed \cii in $z>6$ galaxies \citep[e.g.][]{Riechers2013,Knudsen2016,Venemans2016,Decarli2017,Stanley2019,Venemans2020} as the line has the potential to characterise the ISM at this early epoch. A large number of detections have been discovered across a range broad of SFRs \citep[e.g.][]{Capak2015,Willott2015,Knudsen2016, Pentericci2016,Bradac2017,Decarli2017,Matthee2017, Carniani2018, Smit2018,Hashimoto2019,Bakx2020,Bethermin2020,Harikane2020,LeFevre2020,Fujimoto2021,Laporte2021}, but interestingly also a number of non-detections \citep[e.g.][]{Ouchi2013,Gonzalez2014,Ota2014,Maiolino2015,Schaerer2015,Knudsen2016}.  

Because of the faint absolute UV magnitude of sub-$L^*$ galaxies at $z>6$ \citep[e.g.][]{Ono2018}, one would need at least 10 hours of ALMA observing time to detect them in \cii. In this context, and to establish a large sample of sub-$L^*$ galaxies at $z>6$, observations can be done more efficiently by targeting galaxy cluster fields as gravitational lensing will amplify the light from such distant galaxies \citep[e.g.][]{Richard2011,Knudsen2016,Fujimoto2021,Laporte2021}. Even then many galaxies remain undetected, the stacking of the data can enable a significant statistical improvement in the sensitivity \citep[e.g.][]{Fujimoto2018,Fujimoto2019,Stanley2019,Bethermin2020,Carvajal2020,Uzgil2021}.  

In this paper we present a spectral line stacking analysis of the \cii line for a sample of 52 gravitationally lensed galaxies from the ALMA Lensing Cluster Survey (ALCS; Kohno et al. in prep). %\citep[ALCS; ][in prep]{kohno21}. 
The ALCS is a large project observing 33 galaxy clusters using the Atacama Large millimetre/sub-millimetre Array (ALMA). The target sources of this study are in the redshift range $5.9\lesssim z\lesssim6.6$ and have not been individually detected with \cii in the ALCS data.

The structure of this paper is as follows. In Sect. 2 we describe in detail the data as well as the studied sample. Sect. 3 outlines the method and the different tools used to carry out this analysis. In Sect. 4 we present the results before discussing their implications, biases, and newly raised questions in Sect. 5. Throughout the paper we assume $H_0=70$\,km/s, $\Omega_\mathrm{M} =0.3$, and $\Omega_\Lambda=0.7$.

%--------------------------------------------------------------------
\section{Data and sample}

\subsection{ALMA Lensing Cluster Survey}

The ALCS is an large ALMA programme accepted in cycle 6 (Project ID: 2018.1.00035.L; PI: K. Kohno). The programme observed 33 massive galaxy clusters, 16 from RELICS \citep{Coe2019}, 12 from CLASH \citep{Postman2012}, and 5 from the Frontier Fields survey \citep{Lotz2017}, for a total of 88\,arcmin$^2$ observed in band 6 (i.e. at a wavelength of $\sim\,$1.2\,mm). Observations were carried out between December 2018 and December 2019 (cycles 6 and 7) in compact array configurations C43-1 and C43-2. The entire bandwidth of the survey spans a total of 15\,GHz split over two spectral ranges: 250.0--257.5\,GHz (tuning 1) and 265.0--272.5\,GHz (tuning 2)\footnote{ALCS data are combined with existing ALMA data when available, notably the ALMA Frontier Fields Survey used in this analysis: Project ID: 2013.1.00999.S, PI: Bauer and Project ID; 2015.1.01425.S: PI: Bauer}. As ALCS utilises two spectral frequency setups, the survey provides a frequency coverage twice the size of a single setup, and this means a wider redshift coverage. The data were reduced and calibrated using the Common Astronomy Software Applications (\textsc{CASA}) package version 5.4.0  \citep{CASA}. Data cubes were imaged using the \textsc{CASA} task \textsc{TCLEAN} down to a level of 3$\sigma$ with a channel size of 60\kms. Throughout this paper, we use natural-weighted, primary-beam-corrected, and {\it uv}-tapered maps, with a tapering parameter of 2\,arcsec (pixel size of 0.16 arcsec). We use {\it uv}-tapered maps so that all the stacked sources share a similar beam size, and hence a similar spatial resolution, which is essential for stacking homogeneity. However, because high-redshift sources may be small compared to the tapered beam size and for completeness, we also performed our stacking analyses on cubes in their native resolution ($\sim$ 1\,arcsec; see Fig. \ref{CII_nativeRes} in the appendix). A full description of the survey and of its main objectives will be presented in a separate paper (Kohno et al. in prep). The main characteristics of the data cubes used in this analysis can be seen in Table \ref{tab.cubeInfo}.

\begin{table*}
\caption{General information on the ALMA data.}
\label{tab.cubeInfo}
\centering    
 \begin{tabular}{lccccc} 
 \hline\hline
ID$^{(a)}$ & RMS tune 1$^{(b)}$ & RMS tune 2$^{(c)}$ & RMS continuum image$^{(d)}$ \\
 & [mJy/beam] & [mJy/beam] & [mJy/beam] \\ 
 \hline
Abell2744 & 2.6 & 2.1 & 0.17 \\   
MACSJ0416.1-2403 &  2.8 &  3.5 & 0.14 \\
AbellS1063 & 2.1 &  2.6 &  0.17 \\
MACSJ1149.5+2223 &  2.0 &  2.4 & 0.16 \\
MACS1206.2-0847 &  2.4 &  3.0 & 0.15 \\
RXJ1347-1145 &  2.4 & 3.0 & 0.15 \\
MACS0329.7-0211 &  2.6 & 3.3 & 0.18 \\
MACS1115.9+0129 &  2.9 & 3.6 & 0.20 \\
Abell383 & 2.8 & 3.5 & 0.19 \\
\hline
\end{tabular}
\tablefoot{${(a)}$ Cluster name; ${(b)}$ RMS in 60\,\kms channels in the first tuning (250.0-257.5\,GHz, unit: mJy/beam, 147 channels in total); ${(c)}$ RMS in 60\,\kms channels in the second tuning (265.0-272.5\,GHz, unit: mJy/beam, 139 channels in total); ${(d)}$ RMS in continuum image (mJy/beam).}
\end{table*}

\subsection{Sample}

\subsubsection{Sample selection} \label{sub:sampleSelect}

To perform our spectral stacking analysis we select sources with redshift $5.97\lesssim z\lesssim6.17$ or $6.38\lesssim z\lesssim6.60$, that is sources whose potential \cii emission lines would fall in the observed bandwidth. Because redshift uncertainties have a drastic impact on the performance of spectral stacking analyses \citep{Jolly2020}, only sources with spectroscopic redshifts or high precision photometric redshifts ($\sigma_z<0.02$, corresponding to $\sigma_v \sim$ 800\,km\,s$^{-1}$) were selected. This ensures that line peaks are well known and lines are stacked aligned. Sources are extracted from (ordered by number of sources): The MUSE GTO (MGTO) programme catalogue \citep{Richard2020}, the ASTRODEEP catalogue \citep{ASTRODEEP1,ASTRODEEP2,ASTRODEEP3}, the Hubble Frontier Field Deep Space (HFFDS) catalogue \citep{HFFDSpaper}, and the VLT/MUSE observations catalogue \citep{Caminha2019}.

In addition, the regions of interest were queried using AstroQuery \citep{Astroquery2019}, leading to the addition of two sources from the NASA/IPAC Extragalactic Database (NED\footnote{http://nedwww.ipac.caltech.edu/}). We also manually added the two images of a lensed source in Abell 383 \citep[][]{Richard2011,Knudsen2016}. HST catalogues are corrected for astrometry offset with respect to ALMA maps following \citet{Franco2018}.  

To avoid duplicates between different catalogues, sources are removed if they are closer than 1 arcsec from one another. Priority is given to catalogues with higher redshift precision. 

For homogeneity, and to push the limit of observability towards fainter sources, we only select sources that are not individually detected in our ALMA data. After these considerations we selected a total of 52 sources (of which 23 are multiple images of same sources): 32 from MGTO,  10 from ASTRODEEP, four from HFFDS, two from VLT/MUSE, two from \citet{Richard2011,Knudsen2016}, and two from our AstroQuery query. The main characteristics of the sample are presented in Tables \ref{table:sample} and \ref{table:physics}.

From this sample, we assemble four different sub-samples as follows. As shown in \citet{Jolly2020} redshift uncertainty lead to a reduced efficiency of spectral stacking analyses. We hence perform our analysis concurrently on one sub-sample composed of only 36 sources with secure, spectroscopically determined redshifts; that is all sources with photometric redshifts, or sources with low spectroscopic redshift confidence are removed from this sub-sample (see Table \ref{table:sample}).

We also split sources by their SFR, creating a sub-sample consisting solely of sources whose SFRs are greater than the median SFR of the full sample (2.05\,M$_\odot$\,yr$^{-1}$). Finally, we create a sub-sample composed of the sources present in both the high redshift precision and high-SFR sub-samples. 

In the rest of this paper, the different samples are referred to as: 'full sample', 'good-$z$ sub-sample', 'high-SFR sub-sample', and 'SFR-$z$ sub-sample'.

\subsubsection{Sample properties and SED fitting} \label{sampleProperties}

The $z\sim 6$ galaxies used in this study are located behind nine clusters from the Frontier Fields and CLASH surveys, for which deep public NIR data are available \footnote{HST Frontier Fields : https://archive.stsci.edu/prepds/frontier/} \footnote{CLASH : https://www.stsci.edu/~postman/CLASH/ }. Data from three clusters have been analysed as part of the ASTRODEEP project \citep{Merlin2016}, namely Abell2744, MACSJ0416.1-2403 and MACSJ1149.5+2223, and robust catalogues have been released. In addition, Abell2744 and MACSJ0416.1-2403 as well as MACS1206.2-0847, MACS0329.7-0211 and RXJ1347-1145 were analysed as part of the MUSE GTO programme, and complete catalogues were made available through their latest data-release \citep{Richard2020}.  
In the following, we use the catalogues from the ASTRODEEP project and the MUSE-GTO data release for these six clusters, and we use our own catalogues for the remaining. 

We built our catalogues using SExtractor v2.19.5 \citep{SExtractor} in double image mode, using a sum of WFC3 data as a detection picture. We extracted sources on PSF-matched data, obtained after running TinyTim \citep{TinyTim}, using extraction parameters adapted to identify faint sources :  DETECT\_MINAREA\,=\,5\,pixels,  -DETECT\_THRESH\,=\,1.5$\sigma$ and -DEBLEND\_MINCOUNT\,=\,0.00001. The number of extracted sources per cluster ranges from 4,942 (MACS1115.9+0129) to 9,079 (RXJ1347-1145) for the CLASH clusters and goes up to 20,996 for the Frontier Field cluster AS1063 because of the depth difference between the surveys (F105W$\sim$27.5AB vs. $\sim$30.3 AB at 2$\sigma$ for the CLASH and Frontier Fields data, respectively). We extracted the photometry on IRAC 3.6\,$\mu$m and 4.5\,$\mu$m images in a 1.2\arcsec radius aperture centred at the F160W position for isolated objects. Three objects in our sample required more careful IRAC photometry extraction (namely A383-19, RXJ1347-6, RXJ1347-8). For those objects, we used GALFIT \citep{GALFIT} with the following assumptions for the input parameters: We fixed the candidate (and the neighbouring object) positions to the centroids of the F160W/HST image; we adopted Serfic galaxy profiles; the flux, the axes' ratio, and the half-light radius were left as free parameters. The error bars were computed from the noise measured directly on the original IRAC images.  In case of non-detection we used the 2$\sigma$ upper limits in the following. Among the 52 galaxies in our sample, 33 are identified in our catalogues. A visual inspection at the position of the remaining objects show no UV counterpart, suggesting they have a high Ly-$\alpha$ equivalent width (EW).

Because all galaxies in this sample are located behind lensing clusters, one needs to correct the photometry for lensing magnification. For targets in the Frontier Fields data, we used the online magnification calculator\footnote{https://archive.stsci.edu/prepds/frontier/lensmodels/\#magcalc}, and chose the magnification factors computed with the CATS group models (\citealt{Jauzac2014,Richard2014}). For the CLASH clusters, we used the updated models from \citet{Zitrin2015}. 

To estimate the physical properties, such as the stellar mass, age, and SFR,  of the galaxies discussed in this paper, we ran BAGPIPES \citep{BAGPIPES} with four different star formation histories (SFHs): a burst, a constant, an exponential, and a combination of a young burst with a constant SFH. The IMF used in BAGPIPES is from \citet{KroupaandBoily2002}. For all models, the parameter spaces allowed for the age of the component, defined as the beginning of the star formation, was between 0 and 1.0 Gyr, for the mass formed between $10^6$ and $10^{12}$ M$_{\odot}$, for the reddening between $A_V=0.0$, and 1.0\,mag and we fixed the ionisation parameter at $\log U=-2$ (see for example \citealt{Stark2015}). To determine the SFH that gives the best fit of the lensed-corrected SED, we applied the Bayesian information criterion (BIC) accounting for the difference in the number of parameters used in each SFH. For most of the objects, the best fit is obtained either with a constant or burst SFH, with a similar fit quality in most cases. 

Among our sample, $\approx80\%$ of our targets show consistent stellar masses for either a burst or constant SFH, for the remaining $\approx 20\%$, we used the stellar mass obtained from the best fit. The SFR is computed using three methods : (i) taking the stellar mass and average age of the source, which reflects the mean SFR over the age of the galaxy, (ii) using the UV luminosity \citep{Kennicutt1998} and corrected for dust attenuation, and (iii) taking the instantaneous SFR obtained from the SFH, which gives the SFR at the observed redshift. The last method is inconsistent with the two first methods and strongly depends on the age of the burst and even more strongly on the current star formation. For the majority of sources in our sample, the SFR averaged over the age of the galaxy is $<$10\,M$_{\odot}$/yr. The SFRs quoted in the rest of this paper were computed using the UV luminosity and corrected for dust attenuation (method (ii)). The attenuation of the UV light by the dust is relatively modest, with an averaged value of $A_v$=0.20$\pm$0.11 mag.

For sources not detected we used the information from the literature when available. 19 sources, mostly Lyman-alpha emitters from the MUSE GTO catalogues, are not detected in any of the HST bands and we could hence not derive reliable SFRs. This should not impact our results much since the SFRs of the rest of the sample are homogeneously distributed and that we only use the mean and median SFR of the sample, and not individual values. 

\begin{figure}
   \centering
   \includegraphics[width=\hsize]{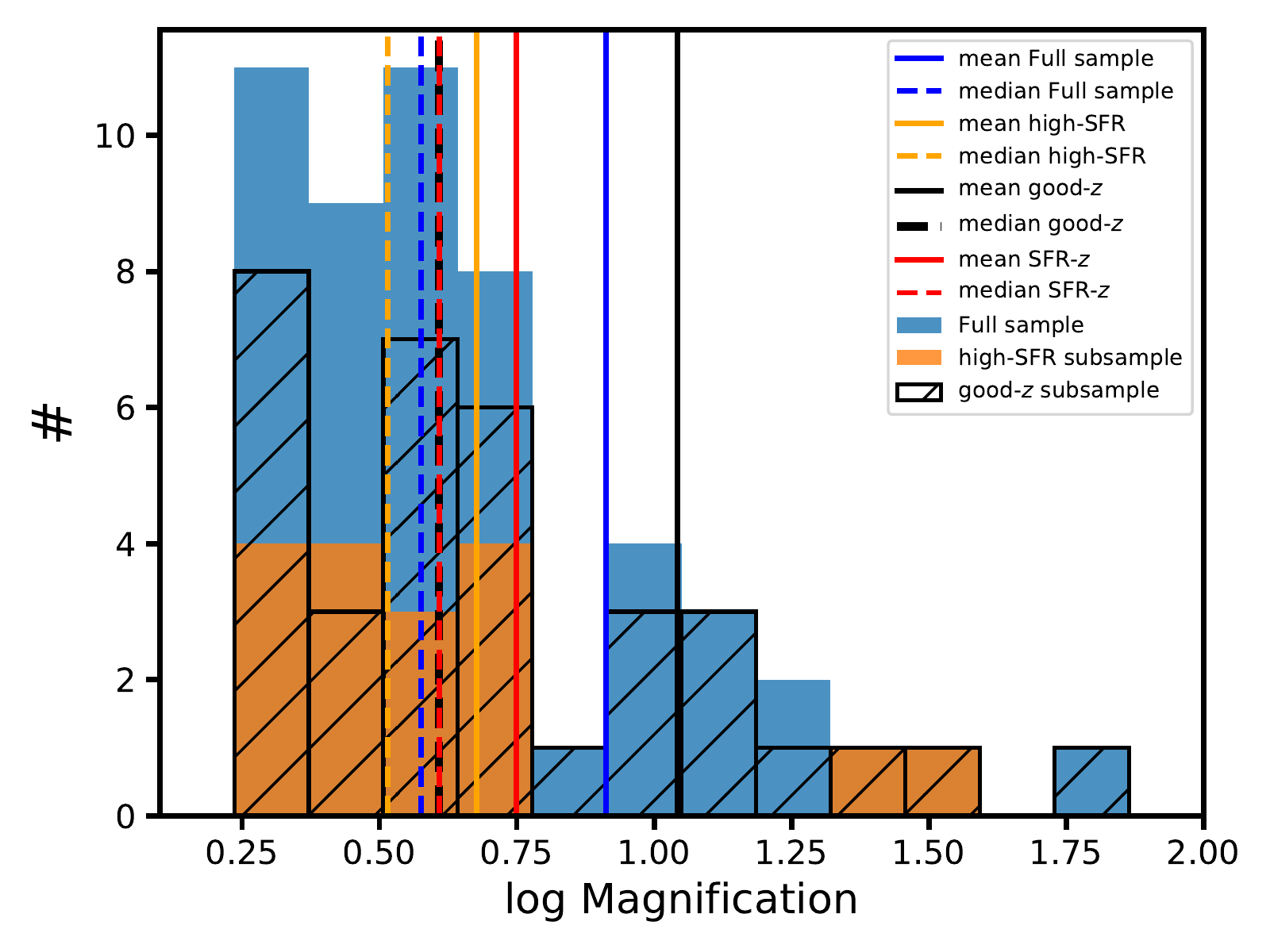}
      \caption{Distribution of the magnifications of the sources across the full-sample (blue), high-SFR sub-sample (orange), and good-$z$ sub-sample (black striped). Plain and dashed straight lines indicate the corresponding mean and median values, respectively, for each sample.}
         \label{MagDistrib}
   \end{figure}

\begin{figure}
   \centering
   \includegraphics[width=\hsize]{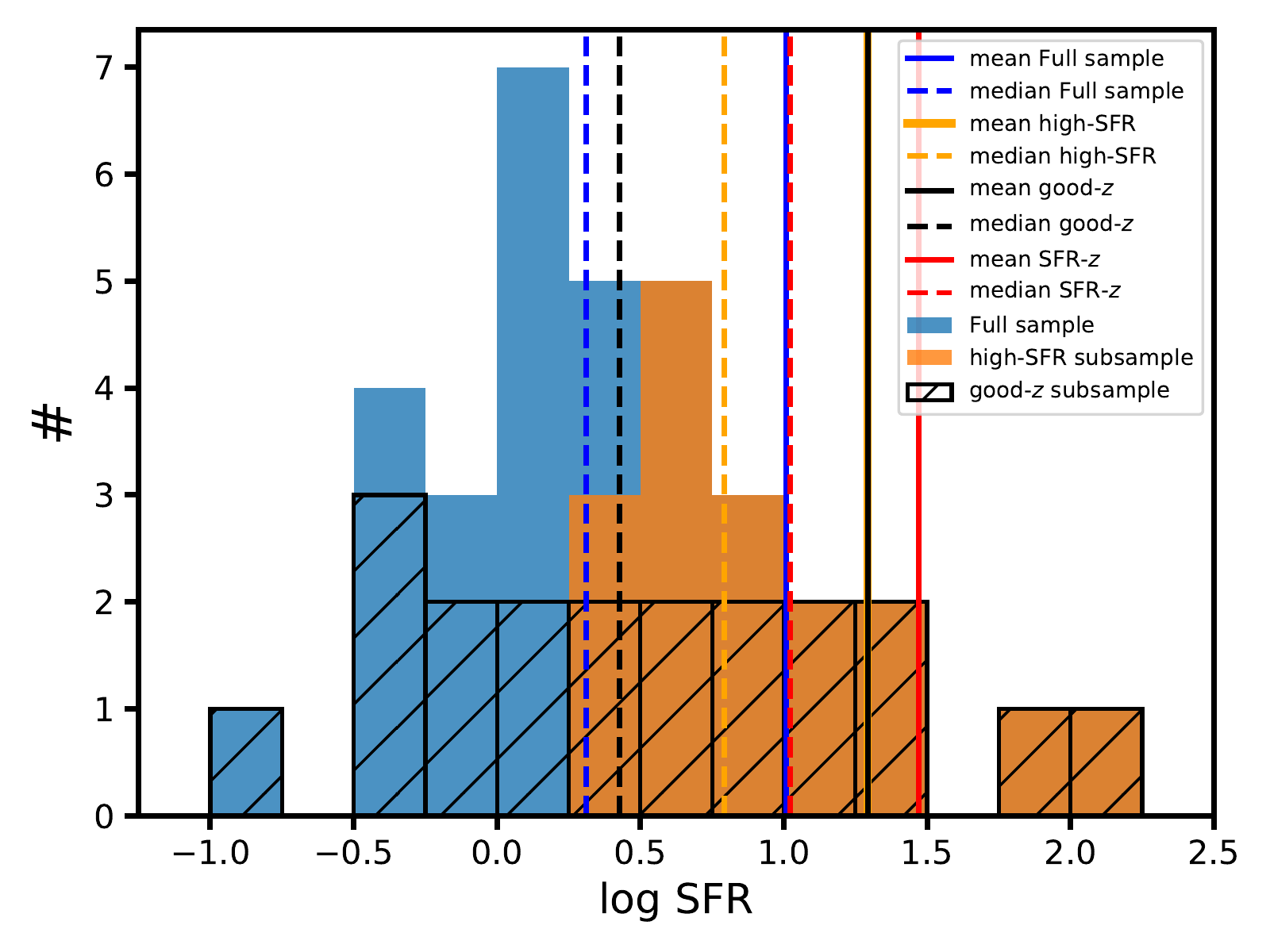}
      \caption{Distribution of the SFR of the sources across the full-sample (blue), high-SFR sub-sample (orange), and good-$z$ sub-sample (black striped). Plain and dashed straight lines indicate the corresponding mean and median values, respectively, for each sample.} 
         \label{SFRDistrib}
   \end{figure}

\begin{figure}
   \centering
   \includegraphics[width=\hsize]{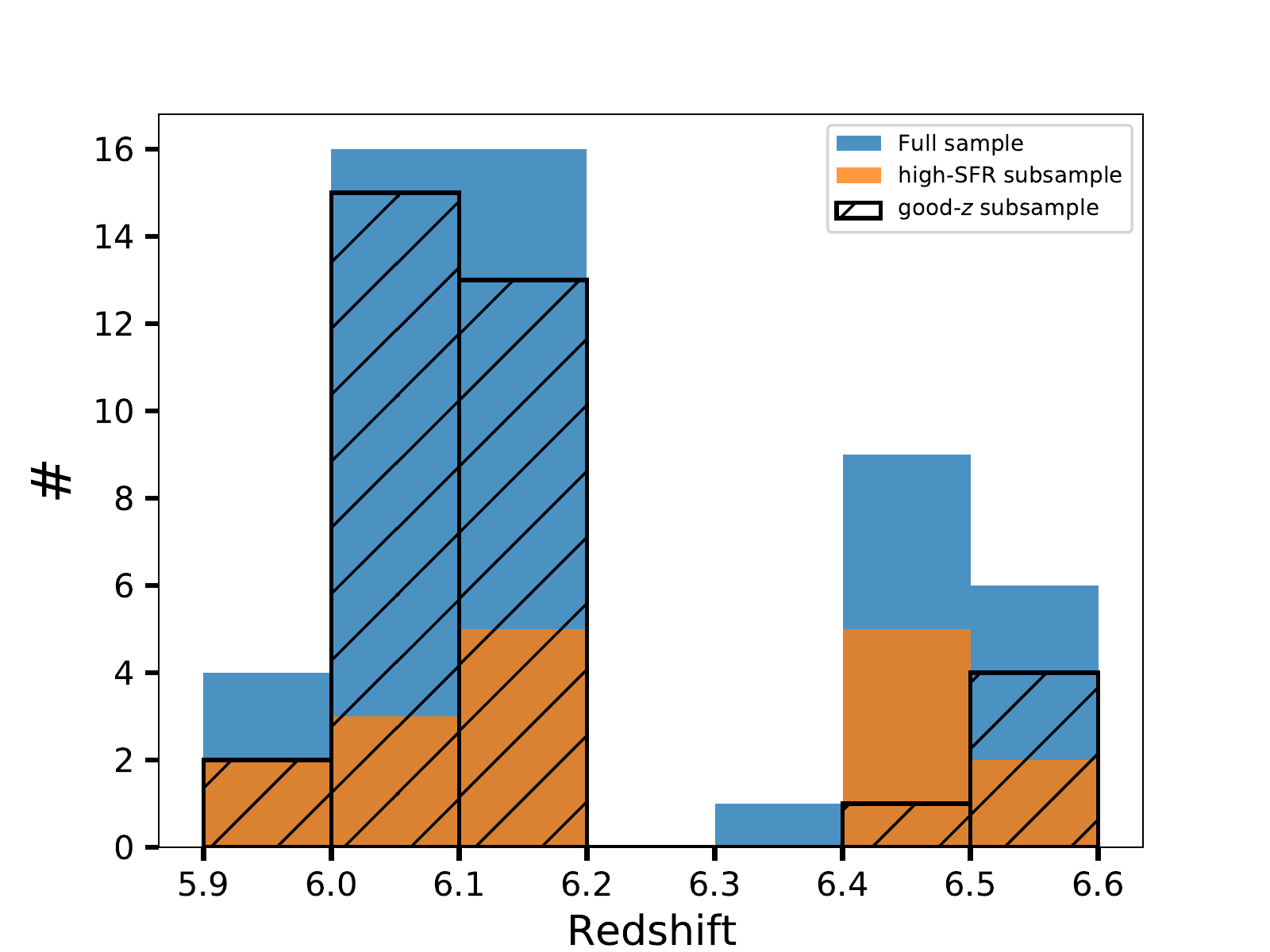}
      \caption{Distribution of the redshift of the sources across the full-sample (blue), high-SFR sub-sample (orange), and good-$z$ sub-sample (black striped).} 
         \label{ZDistrib}
   \end{figure}

The magnification factors of the sources in the full sample range from 1.77 to 72.9 with a median of 3.77 and a weighted-mean\footnote{The weights used to compute the weighted-average magnification and weighted-average SFR are the same weights used in the weighted-average stacking analysis (see Sect. \ref{sec:stacking})} of 8.175 (see Fig. \ref{MagDistrib} and Table \ref{tab.rms}. The magnification corrected SFRs range from 0.17\,M$_\odot$\,yr$^{-1}$ to $\sim\,$125\,M$_\odot$\,yr$^{-1}$, with a weighted-mean\footnotemark[\value{footnote}] of $\sim\,$10.2\,M$_\odot$\,yr$^{-1}$ and a median of $\sim\,$2.0\,M$_\odot$\,yr$^{-1}$, SFR distribution can be seen in Figure \ref{SFRDistrib} and Table \ref{tab.rms}. The redshift distribution across the samples is shown in Figure \ref{ZDistrib}.

In addition to the individual SED fittings described above, we stacked all the best-fit SEDs of the sources obtained with BAGPIPES together (both using mean and median stacking) to recover the SFR$_\mathrm{UV}$. Using these methods we find SFR$_\mathrm{UV, mean}=7.73\pm0.77$\,M$_{\odot}$ and SFR$_\mathrm{UV, median}=2.32\pm0.56$\,M$_{\odot}$.

\section{Stacking method} \label{sec:stacking}

The spectral stacking analysis is performed using \textsc{LineStacker} \citep{Jolly2020}. \textsc{LineStacker} is an open access stacking software developed for stacking interferometric data. It is an ensemble of CASA tasks that allow the stacking of either spectral cubes or extracted spectra. \textsc{LineStacker} comes with a suite of tools, among which are automated weighting routines, which we use in our analysis. 

Sources were stacked pixel to pixel and spectral channel to spectral channel. Both weighted average and median stacking were used. Some sources are located close to the edge of the cubes\footnote{The edges are defined according to the maps compiled in the data (see Kohno et al. in prep)}, making the noise levels of these sources much higher than others. For this reason, we weighted each source proportionally to the inverse of the square of the local RMS. To do so, \textsc{LineStacker}'s automated weighting schemes were used, and local noise was estimated in an area of 12.96\,arcsec$^2$ (81$\times$81 pixels$^2$) around the source. Sources were not masked while computing noise as they were not individually detected in the first place.  

All candidate spectral regions where the corresponding lines would be located for each source were aligned using their redshift and the emission frequency of \cii (1900.5369\,GHz; see Sect. \ref{sec:biases} for a discussion about the impact of the redshift offsets on the stack results). Spatially, sources were stacked using their physical coordinate extracted from the corresponding catalogue. In addition to the spectral stacking analysis, we performed continuum stacking of the same sources in the same way. 

\section{Results} \label{results}

\subsection{\cii stack results} \label{CII_results}

The stacking analyses yielded non-detections for the full sample and all sub-samples. The stacking results are presented as moment-0 maps, integrated over 540\kms\,($-270$\,\kms\  to +270\,\kms) centred on the expected \cii line. \citet{Bethermin2020} and \citet{Schaerer2020} studied a large sample of non-lensed sources through the ALPINE project \citep{Bethermin2020, Faisst2020,LeFevre2020}. Their sample at $z>5$ shows an average SFR of $\sim$\,13\,M$_\odot$\,yr$^{-1}$ for the undetected sources and $\sim$\,82\,M$_\odot$\,yr$^{-1}$ for sources with $9.00< {\rm log}(L_{\rm [CII]} /{\rm L}_\odot)<9.33$, which is either of the order of, or greater than, our sample. Their sample shows a \cii linewidth distribution with a median FWHM of $\sim$252\kms. We decided to integrate over roughly two times their median linewidth in the moment-0 maps, to include lines with potentially larger width as well as to account for artificial increase in the line width that would arise from stacking lines slightly off-centre. Indeed, high uncertainty on the redshift as well as systematic offsets between redshifts derived from Lyman-$\alpha$ or \cii, may result in line emission being stacked with a small velocity shift (the ALPINE survey identified a median offset of 184$^{+201}_{-215}$\kms between Lyman-$\alpha$ and \cii redshift measurement \citep{Faisst2020}; this is discussed further in Section \ref{sec:biases}).

\begin{figure*}
   \centering
    \includegraphics[width=0.9\columnwidth]{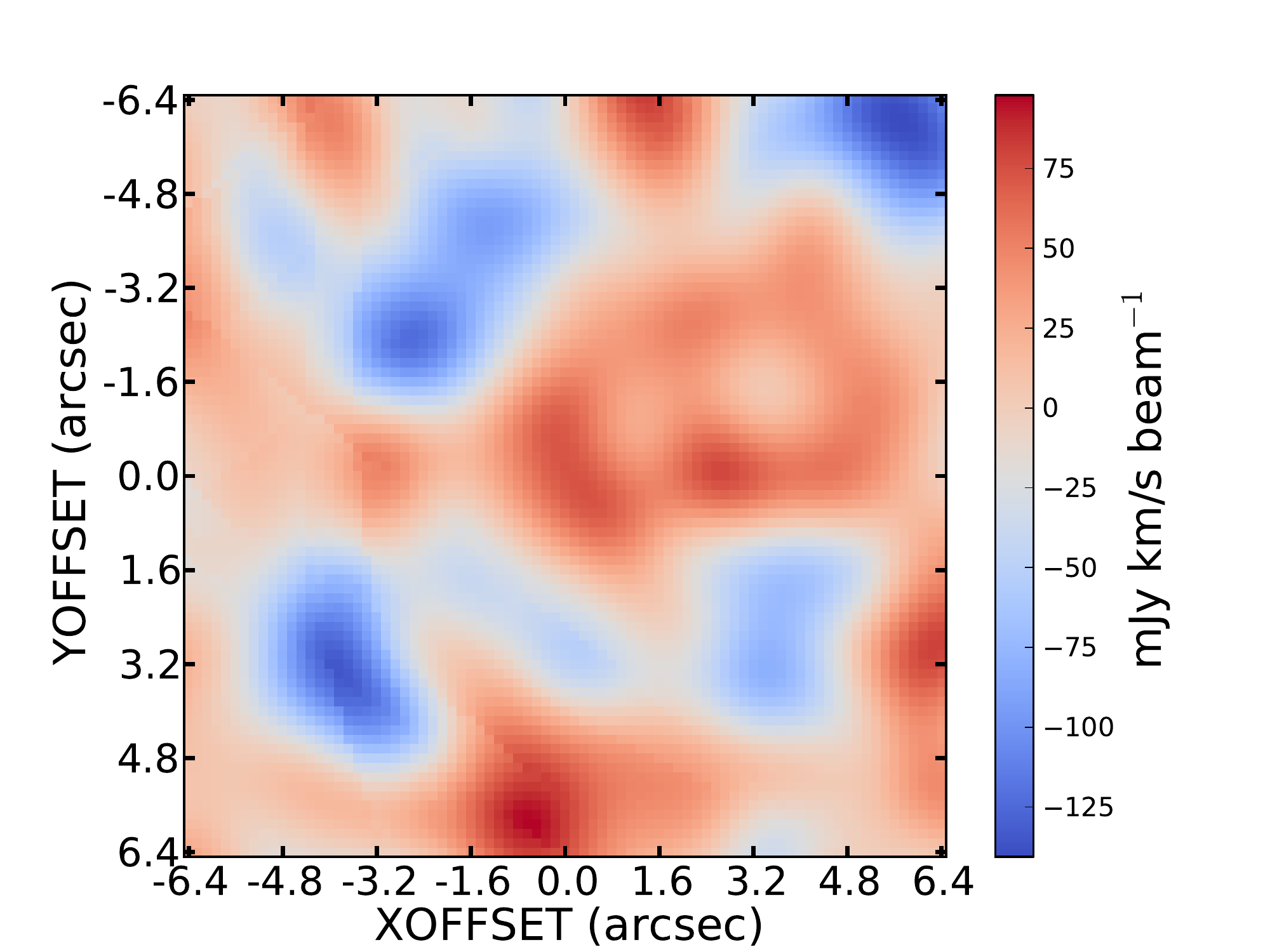}   
    \includegraphics[width=0.9\columnwidth]{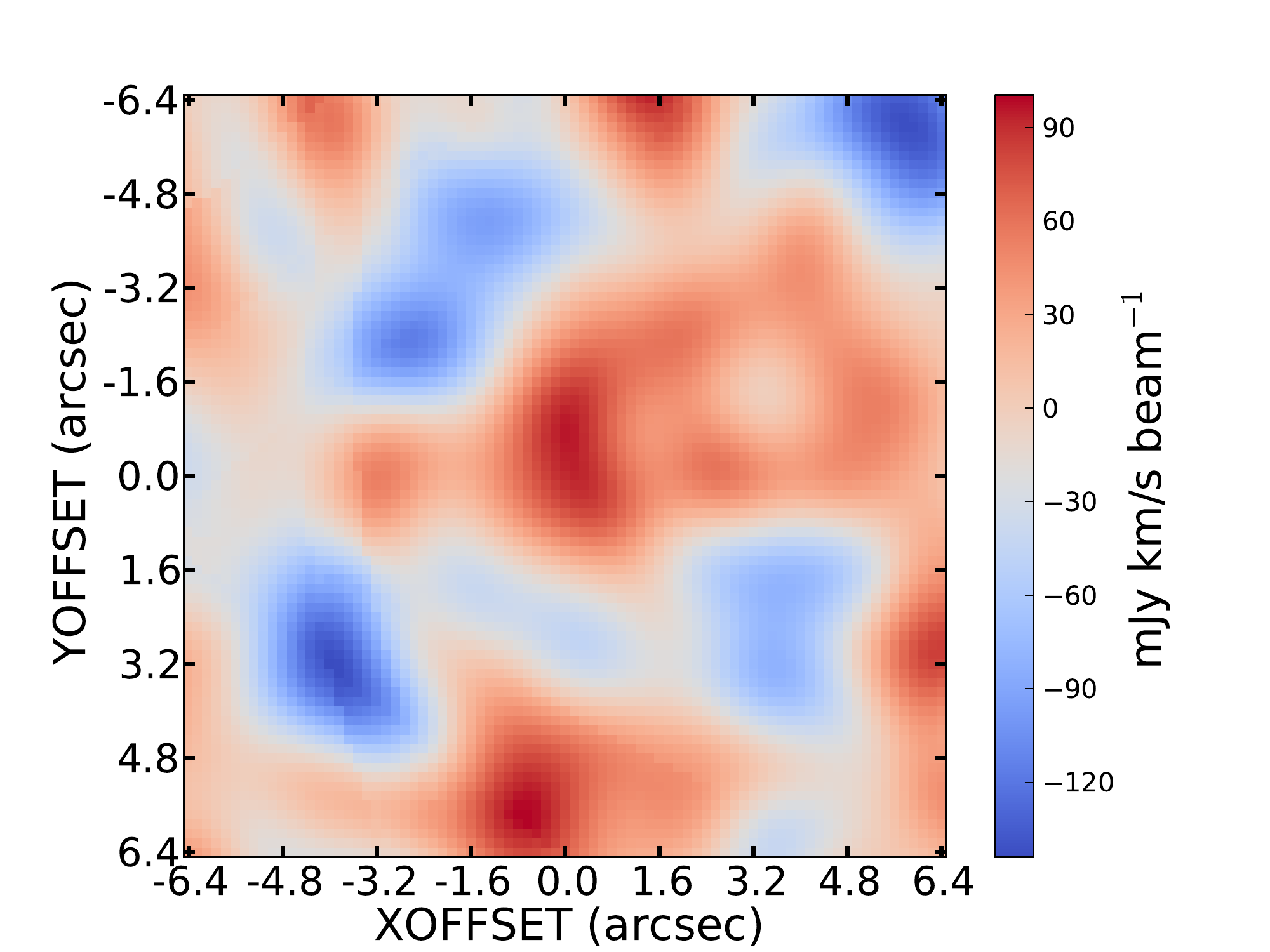}
    \includegraphics[width=0.9\columnwidth]{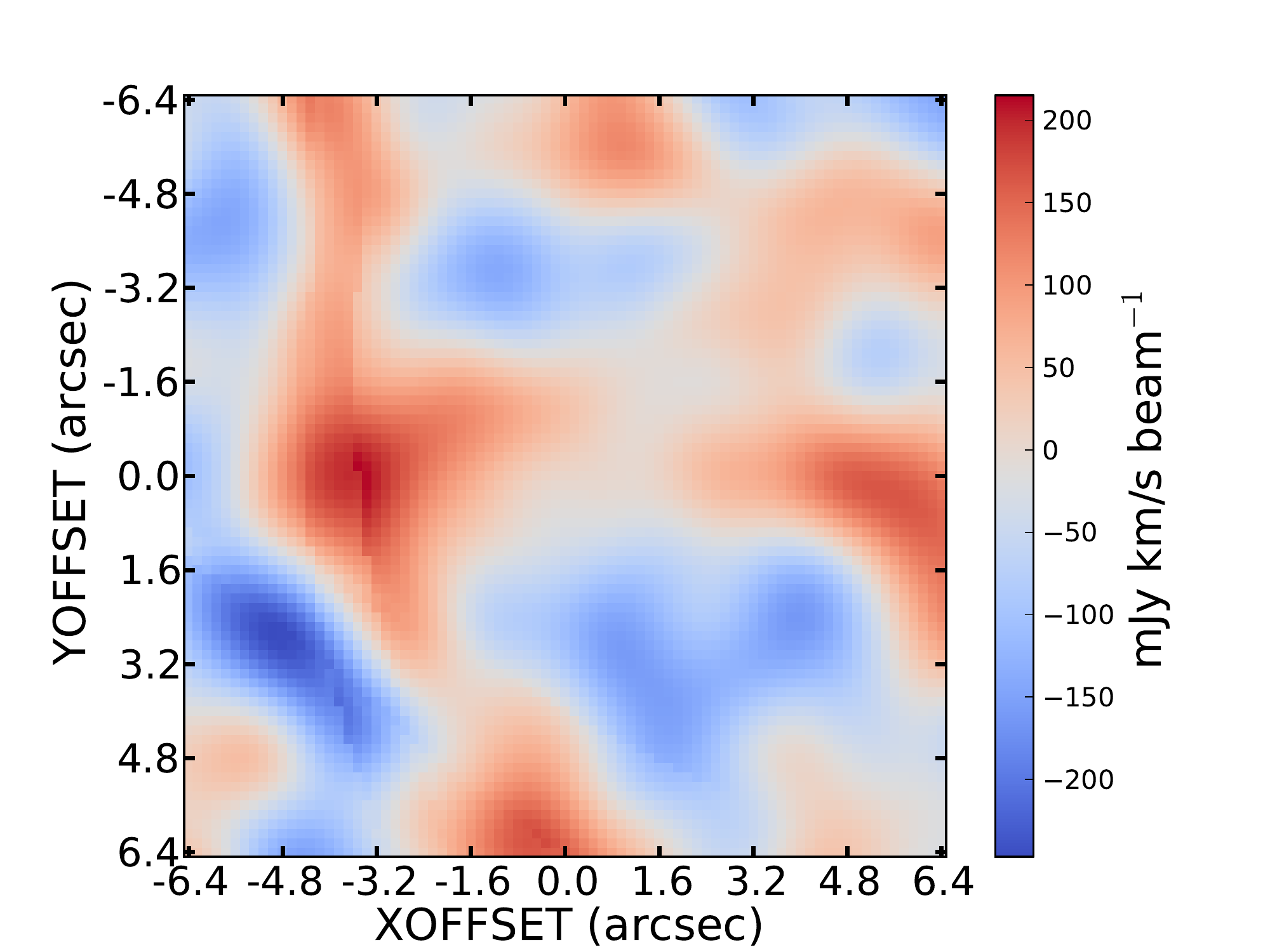}   
    \includegraphics[width=0.9\columnwidth]{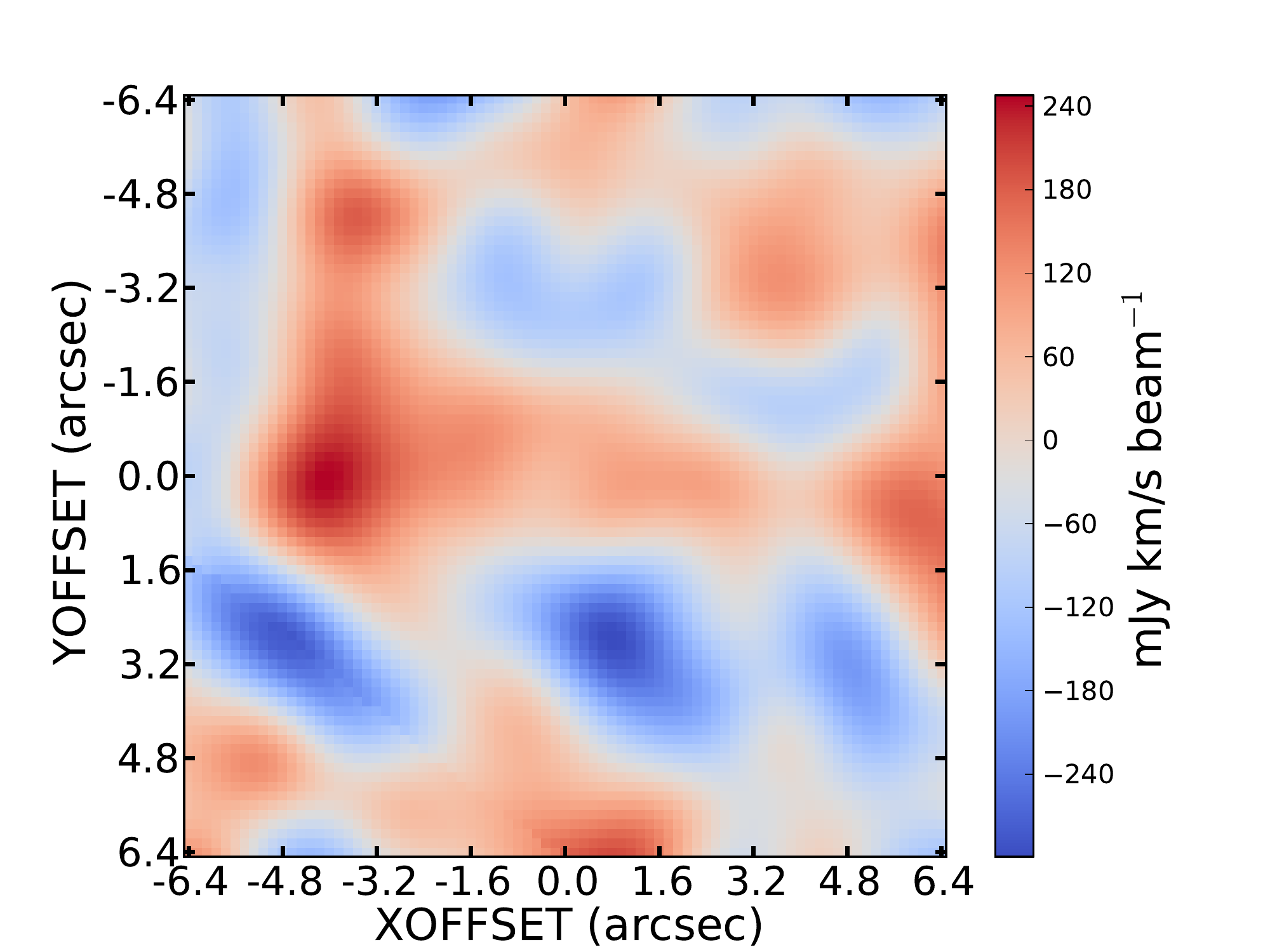}
      \caption{Velocity integrated flux maps of the weighted-average spectral stacks, obtained by collapsing a channel width of 540\,\kms centred on the stack centre. The artificial arcs featured on the sides of the figure are due to stacked sources located close to the edge of the cubes. From left to right and top to bottom, full-sample, good-$z$ sub-sample, high-SFR sub-sample and SFR+$z$ sub-sample.}
         \label{CIIMean_combined}
   \end{figure*}

\begin{figure*}
   \centering
    \includegraphics[width=0.9\columnwidth]{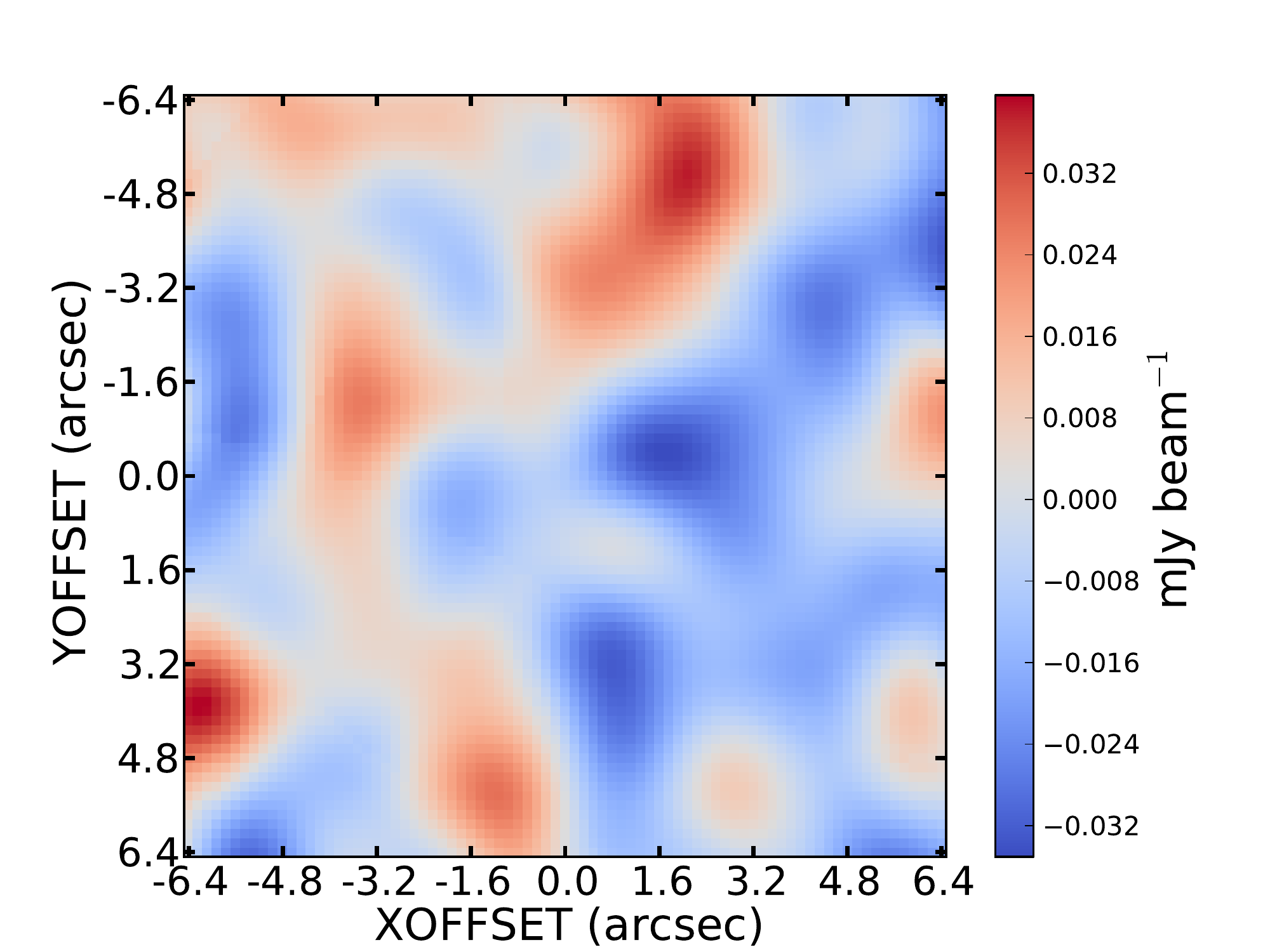}   
    \includegraphics[width=0.9\columnwidth]{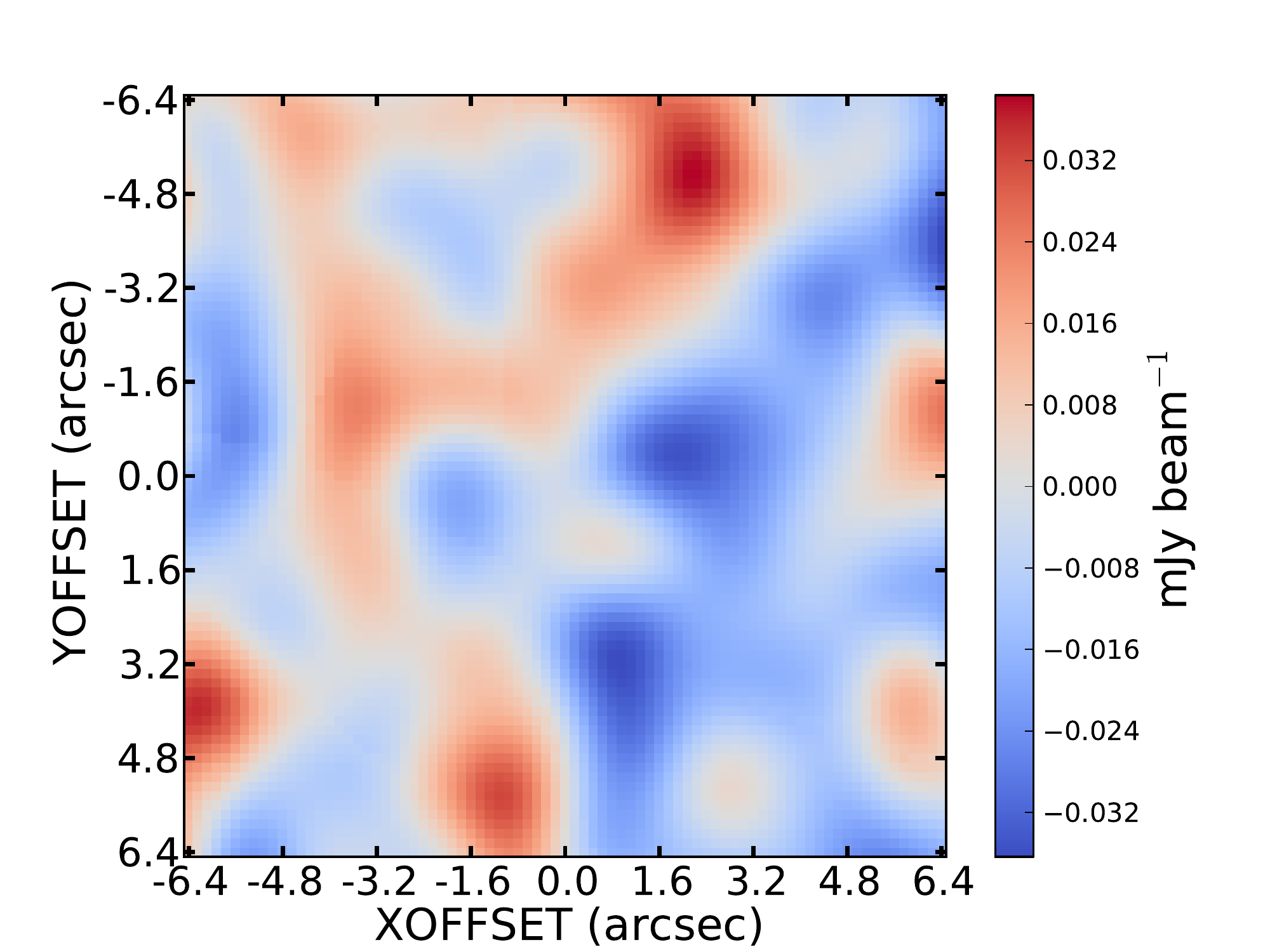}
    \includegraphics[width=0.9\columnwidth]{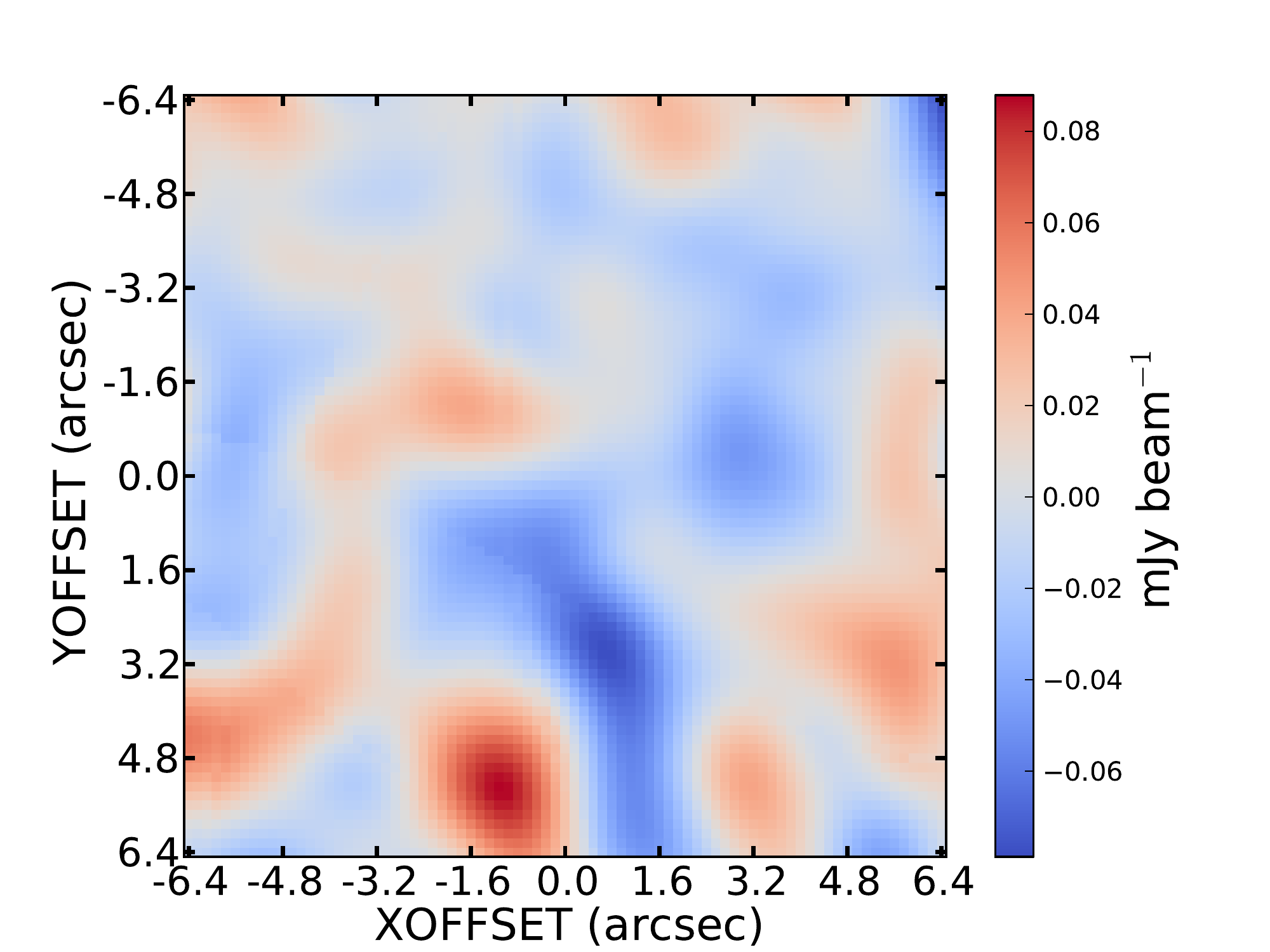}   
    \includegraphics[width=0.9\columnwidth]{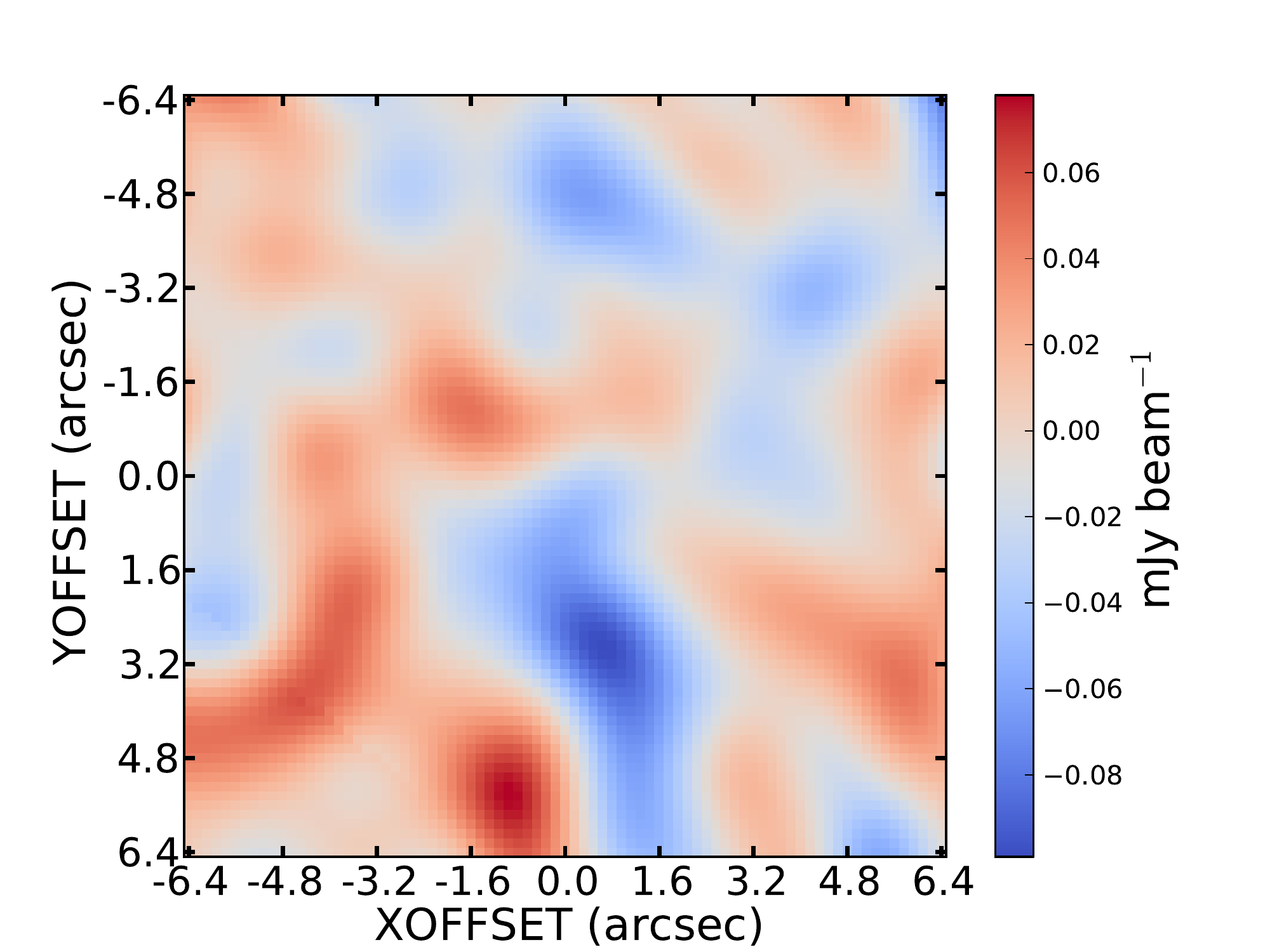}
      \caption{ 
      Continuum maps of the weighted-average continuum stacks. The artificial arcs featured on the sides of the figure are due to stacked sources located close to the edge of the cubes. From left to right and top to bottom, full-sample, Good-$z$ sub-sample, High-SFR sub-sample and SFR+$z$ sub-sample.)
              }
         \label{ContinuumMean_combined}
   \end{figure*}

\begin{figure}
   \centering
   \includegraphics[width=\hsize]{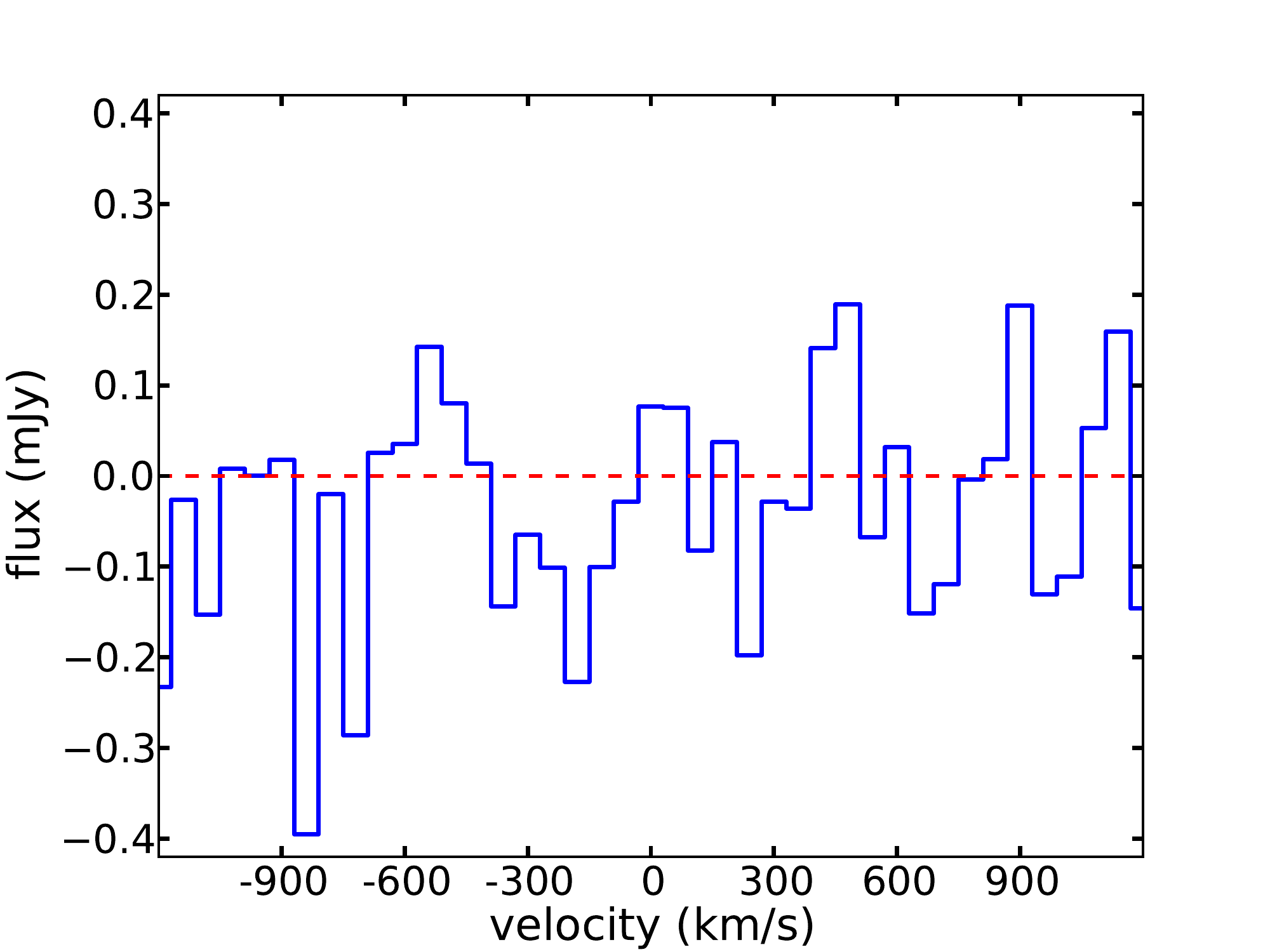}
      \caption{Spectrum extracted from a circular region of radius=1.12 arcsec (7 pixels) centred on the weighted-average stacked cube of the entire sample.}
         \label{CIIMean}
   \end{figure}

Integrated moment-0 weighted-average stacked maps of the samples can be seen in Figure \ref{CIIMean_combined}. Similarly moment-0 median stacked maps of the same samples can be seen in Figure \ref{CIIMedian_combined} in the appendix. In addition to the moment maps we show continuum stack counterpart maps in Figure \ref{ContinuumMean_combined} for the weighted-average stacks and Figure \ref{ContinuumMedian_combined} for the median stack, in appendix. Spectra, extracted from a circular region of radius 1.12\,arcsec (7 pixels) corresponding to the stacked beam of the tapered data, and centred on the stacked cube centre, of both the weighted-average and median stacking analyses of the full sample are shown in Figures \ref{CIIMean} and \ref{CIIMedian}. 

From the absence of detection we derive a 3-$\sigma$ upper limit on \lcii\, similarly to \citet{Schaerer2020} and \citet{Bethermin2020}. \lcii\, is computed using the formula provided in \citet{SolomonVandenBout2005}:
\begin{equation}
    L_{\rm [CII]}=1.04\times10^{-3}\times I_{\rm [CII]}D_L^2 \nu_{obs}\, L_\odot , 
\end{equation}
where $I_{\rm [CII]}$ is the velocity integrated line intensity for \cii in Jy km\,s$^{-1}$ (corresponding here to 3$\,\sigma$ of the stacked moment-0 map), $D_L$ is the luminosity distance in Mpc, and $\nu_{obs}$ is the observed frequency in GHz. The derived \cii line luminosity is then corrected for the weighted mean magnification of the studied sample when performing weighted-mean stacking, or the median magnification when performing median stacking.

\begin{figure}
   \centering
   \includegraphics[width=\hsize]{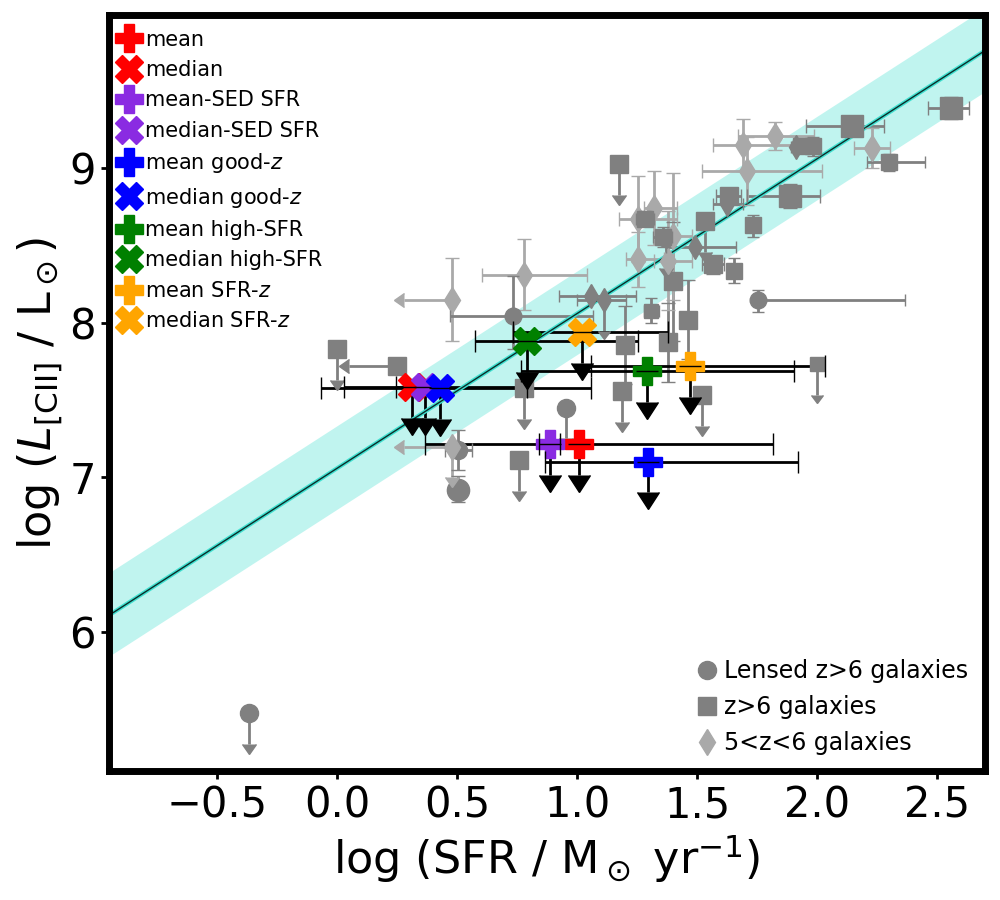}
      \caption{Stacking results, upper limit \lcii\ vs.\ SFR. The error bars of the data obtained in this study are the standard deviation in the case of the mean stacking analyses, and the median absolute deviation for 8.the median stacking analyses (for the SFR obtained from mean and median SED the errors are the errors extracted from the SED fitting). Other data points are extracted from the following studies: \citet{Ouchi2013,Ota2014,Gonzalez2014,Capak2015,Maiolino2015,Schaerer2015,Willott2015,Knudsen2016, Pentericci2016,Bradac2017, Decarli2017,Matthee2017,Carniani2018,Smit2018,Hashimoto2019,Bethermin2020,Bakx2020,Harikane2020,Fujimoto2021}. The cyan line is the relation extracted for low-$z$ starburst galaxies from the \citet{DeLooze2014} study including its $1\sigma$ dispersion.}
         \label{CIISFR}
   \end{figure}

The \lcii\ upper-limits are reported in Table \ref{tab.rms}. The rms obtained from the mean and median stacking methods being comparable, the main difference between the corresponding estimated upper limits on \lcii\ come from the difference between the weighted-average and the median magnification of the samples. Figure \ref{CIISFR} shows the relationship between the derived \lcii\ upper limit and SFR of each sample, obtained through both weighted-average and median stacking. Results from mean stacking analyses are shown with the corresponding mean SFR, results from median stacking analyses with the corresponding median SFR. While upper limits from the median stacking analyses are still compatible with the local relationship, upper limits derived from the mean stacking analyses seem to indicate a lower value of \lcii\ compared to the local relation.

%To draw a relationship between the computed upper limits and the corresponding SFR, we use the weighted-average SFR to compare to weighted-average stacking, and the median SFR to compare to the median stacking analyses, see Figure \ref{CIISFR}. 

Because none of the stacking analyses lead to a detection we do not perform the usual statistical tests designed to probe the significance of the stacking result. Indeed, bootstrapping tests, for example, probe the distribution of the stacked line flux; however, such tests would not serve much purpose if the stacked source is undetected.

We have studied the full sample and the defined sub-samples. It would be possible to subdivide the full sample based on other criteria (e.g. on Lyman-$\alpha$ equivalent width); however, given the clear non-detection we do not pursue further sub-samples. 

One can notice the presence of a stripe pattern on the side of the moment maps in Figure \ref{CIIMean_combined}. These are due to a few sources being close to the edge and some outer pixels of the stacking stamps being empty. Sources are discarded if the cube edge is closer than one beam from the stack centre. It should be noted that sources close to the edge will have intrinsically higher noise levels and will hence carry a lesser weight in the weighted-average stacks.

Extensive photo-dissociation region modelling from \citet{madden20} provides a recipe for utilising \lcii\ to estimate the total molecular gas mass of low-metallicity dwarf galaxies, $M_{\rm H_2} = 10^{2.12} \times [L_{\rm [CII]}]^{0.97}$. Applying this to our results, and with the assumption that the sample sources have sub-solar metallicity (see Section \ref{subsec:models}), we estimate an upper limit on the molecular gas mass, $M_{\rm H_2}$, in the range $ < 1.3\times10^9 - 7.0\times10^9$\,M$_\odot$. This range was derived using the two sample extremes: the $3\sigma$ upper limits derived for (i) the good-$z$ sub-sample, mean stacking analysis and (ii) the SFR+$z$ sub-sample median stacking analysis. Compared to the mean and median stellar masses of $M_{\rm stellar} \sim 1\times10^{9}$ and $2\times 10^{8}$\,M$_\odot$, respectively, these suggest that $M_{\rm H_2}$ is of the order of or higher than $M_{\rm stellar}$, which is in agreement with the molecular gas-to-stellar mass fraction extrapolated from \citet{Walter2020}. 

\citet{Dessauges2020} find an average molecular gas fraction, $f_{\rm molgas}=M_{\rm molgas}$/($M_{\rm stellar}$+$M_{\rm molgas}$), of order $\sim 0.6$ at redshift $4.4<z<5.9$ using a sample of 44 \cii detected non-merger galaxies in the ALPINE survey with a median stellar mass of $\sim10^{9.7}$, indicating a possible flattening of the gas fraction evolution with redshift. To compare to our results, we estimated the mean and median molecular gas fraction for the (sub-)samples using the stellar mass estimate from our SED fitting analysis (we note that stellar masses estimates could not be derived for all sources (see Sect. \ref{sampleProperties}), leading to mean/median masses not exactly representative of the samples as we could only extract $M_{\rm stellar}$ for $\sim 2/3$ of the sources). Our mean results are mostly in good accordance with the average molecular gas fraction derived by \citet{Dessauges2020} as our upper limits indicate $f_{\rm molgas}\lesssim0.5$ (see Table \ref{tab:molgas}). The median results however present higher ratios, of order $\lesssim 0.9$. The large spread between the mean and median results is mainly explained by the difference between the mean and median stellar masses (see Table \ref{tab:molgas}). Indeed, the mean stellar masses are sensible to high-mass outliers, while these have no impact on the median results. As shown in \citet{Dessauges2020}, a dependence is seen between $f_{\rm molgas}$ and  $M_{\rm stellar}$, and the deviation between our own results and the molecular gas fraction cited above can be explained by the difference between the stellar masses probed. In addition, one should note that the $f_{\rm molgas}$ extracted from both our mean and median analyses are mostly consistent with the $f_{\rm molgas}$-$M_{\rm stellar}$ relationship at $z\sim6$ presented in \citet{Dessauges2020}.

\begin{table*}
\caption{ Spectral stacking results. }
\label{tab.rms}
\centering    
 \begin{tabular}{lcccc} 
 \hline\hline
sample name & rms$^{(a)}$ & \lcii$^{(b)}$ & <SFR>$^{(c)}$ & <$\mu$>$^{(d)}$\\%&SFR no corr\\
 & [mJy/beam\,\kms] & [$10^7$\,L$_\odot$] & [M$_\odot\,{\rm yr}^{-1}$] &\\
 \hline
Full sample, mean & 44.2 & <1.65 & 10.22$^{+55.12}_{-7.89}$ &8.17 \\%&8.18\\
Full sample, median & 47.7 & <3.85& 2.05$^{+4.16}_{-0.98}$ &3.77 \\%&4.0\\
\hline
Good $z$ sub-sample, mean & 45.9 & <1.27 & 19.70$^{+63.46}_{-12.37}$ & 11.04 \\%& 11.9\\
Good $z$ sub-sample, median & 50.3 & <3.77 & 2.68$^{+8.75}_{-1.82}$ &4.06 \\%& 5.4\\
\hline
High-SFR sub-sample, mean & 76.7 & <4.91 & 19.57$^{+60.78}_{-13.72}$ &4.76 \\%&12.9\\
High-SFR sub-sample, median & 81.9 & <7.63 & 6.2$^{+11.69}_{-2.44}$ &3.27 \\%& 6.2\\
\hline
SFR+$z$ sub-sample, mean & 96.9 & <5.25 & 29.59$^{+78.10}_{-18.15}$ & 5.62 \\%&16.5\\
SFR+$z$ sub-sample, median & 115.8 & <8.69 & 10.51$^{+13.52}_{-5.13}$ & 4.06 \\%& 8.4\\
\hline
\end{tabular}
\tablefoot{${(a)}$ Rms value in the velocity integrated flux map, obtained by collapsing a channel width of 540\,\kms (nine 60\kms channels); ${(b)}$ Magnification corrected \cii line luminosity computed as discussed in section \ref{results} ; ${(c)}$ Magnification corrected SFR of the sample (see Table \ref{table:physics}): weighted-average SFR of the sources in the sample for weighted-average stacking analyses and median SFR for the median stacking analyses. The errors represent the positive and negative standard deviations for the mean stacking analyses, and the positive and negative median absolute deviations for the median stacking analyses ; ${(d)}$ Magnification of the sample, weighted-average magnification of the sources in the sample for the weighted-average stacking analyses, median magnification for the median stacking analyses.}
\end{table*}

\begin{table}[]
\caption{Upper limit on the molecular gas fraction and corresponding stellar mass for the different sub-samples.}
    \centering
    \begin{tabular}{lcc}
        \hline \hline

        sample name & $f_{\rm molgas}^{(a)}$ &  $M_{\rm stellar}^{(b)}$ \\
         & & (M$_\odot$) \\
        \hline
        Full sample, mean & <0.55 & $10^{9.03}$\\
        Full sample, median & <0.93 & $10^{8.38}$ \\ 
        \hline
        Good-$z$ sub-sample, mean & <0.38 & $10^{9.22}$\\
        Good-$z$ sub-sample, median & <0.92 & $10^{8.40}$\\ 
        \hline
        High-SFR sub-sample, mean & <0.65 & $10^{9.31}$\\
        High-SFR sub-sample, median & <0.95 & $10^{8.53}$\\ 
        \hline
        SFR+$z$ sub-sample, mean & <0.59 & $10^{9.45}$\\
        SFR+$z$ sub-sample, median & <0.95 & $10^{8.56}$\\ 
        \hline
    \end{tabular}
    \tablefoot{$(a)$ Upper limit on the molecular gas fraction, $f_{\rm molgas}=M_{\rm molgas}$/($M_{\rm stellar}$+$M_{\rm molgas}$), derived using the \lcii\,upper limits from the different (sub-)samples; $(b)$ Mean/median stellar mass of each (sub-)sample, extracted from our SED fitting analyses.
    }
    \label{tab:molgas}
\end{table}

\begin{figure}
   \centering
   \includegraphics[width=\hsize]{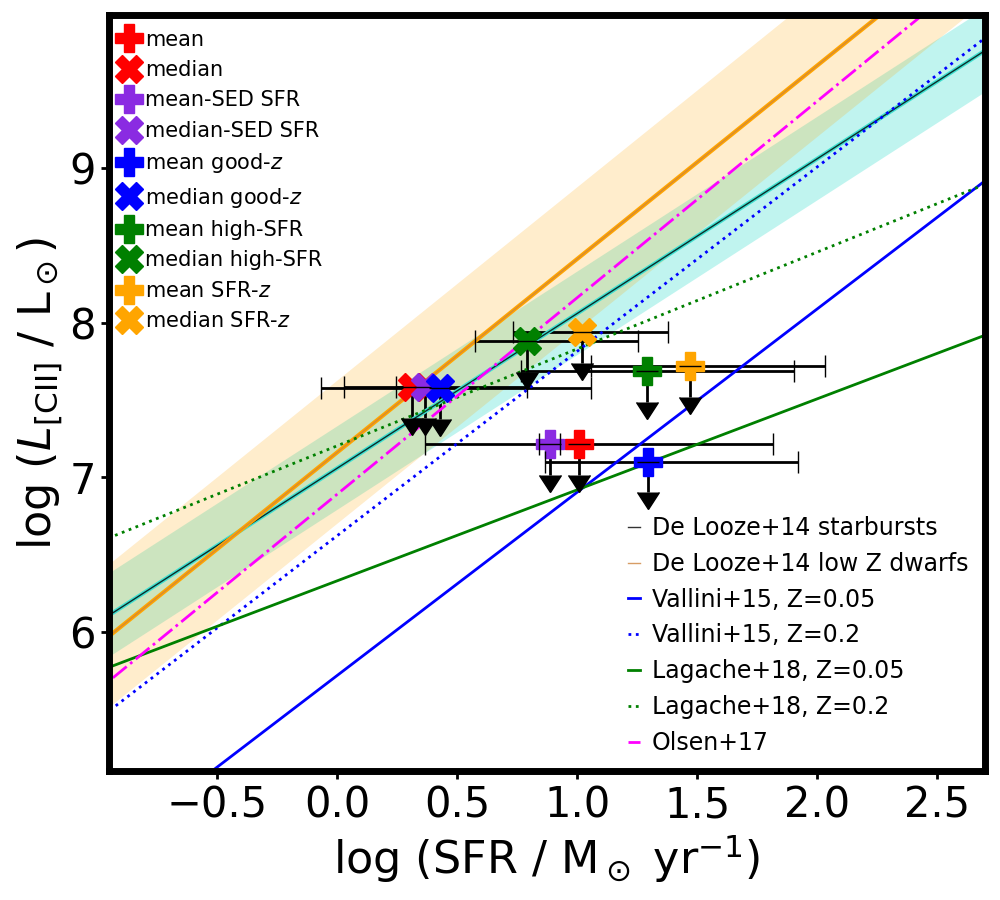}
      \caption{Stacking results, upper limit \lcii vs. SFR compared to several models in the literature. The error bars are the standard deviation in the case of the mean stacking analyses, and the median absolute deviation for the median stacking analyses (for the SFR obtained from mean and median SED the errors are the errors extracted from the SED fitting). Models are extracted from the following studies: \citet{Vallini2015,Olsen2017,Lagache2018}. Relations derived from low-$z$ data come from \citet{DeLooze2014}, shown in cyan and orange, including their $1\sigma$ dispersion. 
              }
         \label{CIISFR_MODELS}
   \end{figure}

\subsection{Continuum}

The stacking analyses of the continuum data yielded non-detections, consistent with the results from the spectral line stacking for \cii. The resulting rms for continuum stack of the full sample and the sub-samples are given in Table~\ref{tab:contstack}.  
We estimate a $3\sigma$ upper limit on the far-infrared luminosity assuming that the far-infrared spectral energy distribution (SED) is well described by a modified blackbody with a temperature of $T=35$\,K and $T=45$\,K, both with dust emissivity $\beta = 1.5$. At the high redshift of the sample, the CMB temperature is $T_\mathrm{CMB}\sim20$\,K, and we adopt the prescription from \citet{daCunha13} to correct for the effect of the CMB. 
We note that single photometric upper limit only provides a rough estimate on $L_{\rm FIR}$, which is subject to a large systematic uncertainty depending on, for example, the choice of temperature or $\beta$.  
The far-infrared luminosity upper limits are used to estimate the dust-obscured SFR, and we assume ${\rm SFR}(L_{\rm FIR}) {\rm [M_\odot\,yr^{-1}]}\sim 1.1\times10^{-10} L_{\rm FIR}[{\rm L_\odot}]$ (for a Kroupa IMF, 0.1-100\,M$_\odot$, and constant star formation over a timescale of 100\,Myr; \citealt{calzetti11}).  We note that different assumptions on the IMF and SFH could lead to a factor of two, or more, change in the estimated values \citep[e.g.][]{rieke09,calzetti11}. 
The estimated upper limit on the $\mathrm{SFR}_\mathrm{IR}$ is $\sim1-6$\,M$_\odot$\,yr$^{-1}$ after correcting for magnification, which is either similar to or below the average rest-frame UV derived SFRs of the samples (see Table\,\ref{tab.rms}). This implies that less than half of the star formation is obscured by dust. It is worth noting that the correction for dust extinction on the UV derived SFR presented in Table \ref{tab.rms}, leads to a reduction of the SFR of $\sim3-6$M$_\odot$\,yr$^{-1}$. While these numbers are not fully representative of the sample (since they are only available for the sources that we could analyse ourselves through SED fitting), the sum of $\mathrm{SFR}_\mathrm{UV}$, not corrected for dust attenuation, and $\mathrm{SFR}_\mathrm{IR}$ is comparable to the original $\mathrm{SFR}_\mathrm{UV}$, after correcting for dust attenuation. 
%
%{\color{magenta}\bf ADD TEXT WITH ESTIMATES ON THE DUST MASSES!}
Similarly, using the continuum upper limit we estimate an upper limit on the dust mass.  We assume a dust opacity of $\kappa_\nu = \kappa_0 (\nu / \nu_0)^\beta$, using $\kappa_0 = 5.1$\,cm$^2$/g  for $\nu_0=1.2$THz \citep[e.g.][]{DraineLi07}, and $\beta=1.5$. This results in upper limits in the range $(0.6-8.1)\times 10^6$\,M$_\odot$.  However, we note that there are significant systematic uncertainties associated with this, and the values could be a factor of five higher depending on the assumptions \citep[e.g.][]{LiDraine01,DraineLi07,Magdis12,casey14,Abergel14}. If we assume $\kappa_0 =0.43$\,cm$^2$/g  for $\nu_0 = 350$\,GHz \citep[e.g.][]{Abergel14}, the upper limits are in the range $(1.2-14.7)\times 10^6$\,M$_\odot$. 
With the large systematic uncertainties, this is consistent with the upper limits on the molecular gas mass, and the factor of $\sim100$ between the two upper limits is consistent with our assumption that the studied galaxies have low metallicity \citep[e.g.][]{Leroy2011}. We note that the continuum emission around rest-frame 158$\mu$m is close to the peak of the dust emission, and it is thus more sensitive to the temperature and total luminosity (probing the SFR) rather than the total dust mass, which is better measured by the emission at the Rayleigh-Jeans tail \citep[e.g.][]{Magdis12,Scoville16}.

\begin{table*}
\caption{Continuum stacking results, using $3\sigma$ upper limits corrected for magnification.}
\label{tab:contstack}
\centering    
 \begin{tabular}{lccccccc} 
 \hline\hline
 & & \multicolumn{3}{c}{$T=35$\,K} & \multicolumn{3}{c}{$T=45$\,K} \\
 sample name & rms$^{(a)}$ & $L_{\rm FIR}$ $^{(b)}$ & SFR$^{(c)}$ & $M_{\rm dust}$\,$^{(d)}$ & $L_{\rm FIR}$ $^{(b)}$ & SFR$^{(c)}$ & $M_{\rm dust}$\,$^{(d)}$ \\
 & [$\mu$Jy/beam] & [$10^{10}$\,L$_\odot$] & [M$_\odot\,{\rm yr}^{-1}$] & [10$^6$\,M$_\odot$] & [$10^{10}$\,L$_\odot$] & [M$_\odot\,{\rm yr}^{-1}$] & [10$^6$\,M$_\odot$] \\

 \hline

Full sample, mean &  15 &  < 0.5 &  < 0.5 &  < 1.7 &   <1.1 &   <1.2 &   <0.9\\
Full sample, median &  17 & <  1.3 &  < 1.4 &  < 4.2 &  < 2.8 &   <3.0 &  < 2.2 \\
\hline
Good $z$ sub-sample, mean & 14 &  < 0.4 &  < 0.4 &  < 1.2 &   <0.8 &   <0.9 &  < 0.6 \\
Good $z$ sub-sample, median &  16 &  < 1.1 &  < 1.2 &  < 3.7 & <  2.5 &  < 2.7 &  < 2.0 \\
\hline
High-SFR sub-sample, mean & 25 &  < 1.5 &  < 1.6 & < 4.8 &   <3.3 &   <3.5 &  < 2.5 \\
High-SFR sub-sample, median & 26 & < 2.3 &  < 2.4 & < 7.6 &  < 5.1 &   <5.5 &   <4.0 \\
\hline
SFR+$z$ sub-sample, mean & 28 &  < 1.4 &  < 1.5 &  <4.8 &   <3.2 &  < 3.4 &  < 2.5  \\
SFR+$z$ sub-sample, median & 35 &  < 2.4 &  < 2.6 & < 8.1 &  < 5.5 & <  5.9 & <  4.3 \\
\hline
\end{tabular}
\tablefoot{ ${(a)}$ rms value in the continuum map ; ${(b)}$ upper limit on far-infrared luminosity corrected for the magnification and the effect of the CMB; ${(c)}$ upper limit on SFR estimated from the $L_{\rm FIR}$; ${(d)}$ upper limit on molecular gas mass estimated from the $L_{\rm [CII]}$.}
\end{table*}

As mentioned in section \ref{CII_results}, one can see the presence of stripe patterns on the continuum stacked maps, even more obvious than on the \cii moment-0 maps (see Figure \ref{ContinuumMean_combined}). While only sources further than one beam from the edge of the cubes are selected in the \cii stacked maps, this edge can can be positioned at a slightly different position on the continuum maps. This is due to both the frequency dependency of the primary and synthesized beam and to the two spectral windows used to construct the continuum maps, which sometimes have different spatial coverage. We however decided to construct the continuum stacked maps as pure counterparts of the \cii stacked maps, and hence included strictly the same sources in both analyses, regardless of whether sources were closer to edge in the continuum maps.

\section{Discussion} \label{sec:Discussion}

\subsection{Potential biases} \label{sec:biases}

While our derived \lcii\,upper limit measurements indicates a possible different behaviour for high-z galaxies in the low-SFR regime compared to their equivalent at low-z, we caution that our upper limit could be underestimated.

The first and probably main bias could come from systematic redshift uncertainties. 
The redshifts of most of the galaxies in our samples are based on the Ly-$\alpha$ lines, which have been shown to exhibit possible systematic offsets compared to the redshift of the \cii line \citep{Capak2015,Faisst2016,Matthee2017,Bethermin2020,Cassata2020,Pahl2020}. \citet{Faisst2020} show a median offset of $184^{+201}_{-215}$\,\kms between the \cii measured redshift and redshift measured from Ly-$\alpha$ in the ALPINE survey. Because the offset redshift is not constant, there is no straightforward way to correct for it while stacking, especially since our stacking analysis leads to no detection \footnote{One could otherwise try to maximise the stacked line flux by randomising redshift offsets as a Monte Carlo process. But with a non detection this would most probably lead to maximising noise peaks.}. However, as shown in \citet{Jolly2020}, redshift uncertainties of $\sim$\,180\,\kms\, lead to an average stacked line amplitude of $\sim70$\% of its maximum value, and an average flux of $\sim90$\% of its maximum value (stacking 30 lines of 400 \kms\, with a S/N $\sim$\,1 pre stacking). Such systematic uncertainties would hence only lead to a modest underestimate of the upper limits derived in our analysis. 
 
Similarly, stacking positions could also be a source of systematic uncertainties. Systematic astrometric shifts in the source catalogues compared to the ALCS data would lead to a poor overlap of stacking positions. This impact could be significant if uncertainties are high compared to the \cii size of the sources. This is, however, unlikely as offsets between HST catalogues and ALMA are well known \citep[e.g.][]{Dunlop2017,Franco2018,Fujimoto2019,Franco2020} and accounted for, and MUSE catalogues are properly aligned \citep{Richard2020}. Nonetheless, a similar effect would be expected from systematic offsets between the \cii regions and the optical regions that are the stacking targets.

As suggested by \citet{Carniani2020} \cii extended emission could also lead to reduced emission and an underestimate of the upper limit. Such an effect could be made more probable from stacking objects elongated by gravitational lensing effects. However, from the average low magnifications (most under 10) as well as the absence of evidence of elongated objects from the {\it HST} NIR images in our sample, we do not expect elongated \cii nor continuum emission.   

Normally a basic assumption for a stacking analysis, is that the sample sources are drawn from the same parent population, thus providing a meaningful sample homogeneity. For this particular study, the sample is constructed from several catalogues of different sensitivity and different selection criteria at optical and near-infrared wavelengths, leading to a heterogeneous sample. The primary selection criteria applied here are position, redshift, and luminosity (i.e. \lcii\ below the threshold for individual detection). 
Our sample consists mostly of Ly-$\alpha$ emitters (all spectroscopically identified sources show Ly-$\alpha$ emission), some of which are also Lyman-break galaxies (12 sources in our sample classify as Lyman-break galaxies based on their SED). 
While it is expected that all normal star-forming galaxies have a \cii line, the strength of this line could be affected by different conditions and properties of each galaxy and its environment (see Sect.~\ref{subsec:models}). 
We note that Ly-$\alpha$ emitters are not necessarily representative of the entire population of normal star-forming galaxies at $z\sim6$ \citep[e.g.][]{Stark2011,Tilvi2014,DeBarros2017,harikane18,Pentericci2018,Kusakabe2020}. It should be noted however, that in the good-$z$ sub-sample nearly all galaxies are selected based on a high Ly-$\alpha$ EW, which is a fairly uniform selection.

However, because our sample is biased towards Ly-$\alpha$ emitters, one can expect the studied galaxies to be dust poor. Indeed, if the observed galaxies were dust rich one would expect scattering in the ISM, leading to lower Ly-$\alpha$ emission. Therefore, one can expect to see a lower far-infrared continuum, as well as lower \cii emission.

When looking at the results presented in Figure \ref{CIISFR} one can see a significant difference between the mean and median analyses. This is not due to a drastic difference in the resulting stacked cubes, but comes essentially from the difference between the mean and median magnification of our sample. It is hence obvious that the individually derived magnifications will have a significant impact on the \lcii-SFR limits derived through our analysis. Many lensing models exist to derive magnification values \citep[e.g.][]{Keeton2011,Jauzac2014,Johnson2014,Ishigaki2015,Merten2015,Zitrin2015,Diego2016,Bradac2017} and our model choices (see section \ref{sampleProperties}) have a sizable impact on our result. However, and unless big systematic differences exist between models, selecting a high number of sources should, on average, reduce the impact of choosing one set of models in place of an other.  

Similarly, the uncertainties on the SFR --while depicting the SFR spread over the studied sample-- highlight the importance of the precision of this value. One can indeed see that the spread of the SFR is quite large in our samples and, because of this, the derived \lcii-SFR limits may end up mostly in accordance with \citet{DeLooze2014} local value, or in total disagreement with it. It is hence important to highlight the impact of SFR derivation as well as its uncertainty. Due to the limited amount of data, and the existence of different models, derived SFR can vary by factors of several for the same source. However, and similar to the uncertainties on magnification, uncertainties on the SFR should tend to even out on average if the number of sources is high enough. Our study should hence be less impacted by SFR uncertainties than single objects studies, but these numbers should still be evaluated with caution.   

Finally it should be noted that some stacked sources are multiple images of the same source, this intrinsically reduces the statistical significance of our results as it diminishes the number of independent objects observed.

\subsection{Comparison to individual detections in high-$z$, low-SFR galaxies }

To probe how much our results may differ from analyses on similar objects we compare it to individual detections in high-$z$, low-SFR galaxies.
\cii has been detected in $z>6$ lensed galaxies \citep{Knudsen2016,Bradac2017,Bakx2020,Fujimoto2021,Laporte2021}, and non-detections have been reported for two $z>8$ galaxies \citep{laporte19}.  The estimated magnification-corrected line luminosities are in the range $L_{\rm [CII]} \sim 10^{7-9}$\,L$_\odot$ for SFRs $\sim\,$3-60\,M$_\odot$\,yr$^{-1}$, which partially overlap the SFR distribution of our sample.
The lowest-SFR, non-lensed detection of \cii is seen towards BDF-3299 \citep[$z\sim7.1$, SFR$\sim6$\,M$_\odot$\,yr$^{-1}$][]{carniani17}. 
The reported detections, along with the two non-detections, are either consistent or below the $L_{\rm [CII]}$-SFR relation found in low-$z$ starburst galaxies \citep{DeLooze2014}, as shown in Figure \ref{CIISFR}.

The rms of the stacked maps in this study are comparable to the reported observed rms values for [CII] detections in the literature. This suggests that if all sources in our sample were to share the properties of the individual detected sources, we would have expected at least a tentative detection in our stacked results.  

The $L_{\rm [CII]}$-SFR relation has a known scatter. However, given the large number of detections (and non-detections) of $z>6$ sources distant from the local \lcii-SFR relation, this would either imply that at $z>6$ this scatter is larger, or that the relation for high-redshift galaxies is offset from the local relation.

The ALPINE survey \citep[e.g. ][]{Bethermin2020,Schaerer2020} is another ALMA Large programme presenting \cii observations of 118 individually selected sources in the redshift range $4.4<z<5.9$. The ALCS \cii redshift coverage is effectively continuing to a higher redshift beyond the ALPINE survey. The SFR range of the ALPINE sample is $\sim 10 -300$\,M$_\odot$\,yr$^{-1}$, which partly overlaps with the upper range of our sample \citep{Bethermin2020}.  A comparison between these two sets of results depends on whether we focus on the results from the median or mean stacking. The mean SFR of the mean stacking has a wide scatter because it is dominated by a few high-SFR estimates. The resulting stacked \cii upper limit is lower than that of the ALPINE survey at a similar SFR. Focusing on the median distribution, we note that our stacked results probe a complementary region of the $L_{\rm [CII]}$-SFR plane and remains mostly compatible with the local relationship.   

\subsection{Is a low $L_{\rm [CII]}$ expected for low-SFR, $z>6$ galaxies?} \label{subsec:models}

The comparison of the $L_{\rm [CII]}$-SFR relation from local galaxies to $z>6$ galaxies is challenging.
A fraction of $z>6$ galaxies lying below the relation, including the upper limits derived from the stacked results, might be explained by the physical conditions combined with the early evolutionary stage.  
Metallicity, the hardness and intensity of the UV radiation field, the gas density distribution, and the star formation history are all aspects to be considered.  

Given the relatively short time after the Big Bang, it is possible that the galaxies targeted in our study will not have had sufficient time to reach a solar level metallicity \citep{Stark2016}. In the local universe, dwarf galaxies are known to have on average lower metallicity \citep[e.g.][]{Cormier2015}, and given the comparable stellar masses of dwarf galaxies and the target sources, it is reasonable to assume that the studied sample have sub-solar metallicity.  
The metallicity has been estimated only for a small number of gravitationally lensed $z>6$ galaxies \citep[e.g. ][]{Stark2015}; however, the detections or non-detections of dust also give an indication of the metallicity \citep[e.g. ][]{watson15,Bakx2020}. Previous modelling of photo-dissociation regions suggest that in lower metallicity-environments the \cii luminosity will be reduced \citep[e.g. ][]{Rollig2006}. Modelling of the \cii luminosity for $z>6$ galaxies also shows that the low-metallicity regions have a lower $L_{\rm [CII]}$ \citep[e.g.][]{Vallini2015,Olsen2017,Lagache2018,Ferrara2019}. In addition, as shown in \citet{Vallini2015}, at $z>4.5$ the CMB temperature becomes comparable to the temperature of the cold neutral medium, strongly attenuating \cii emission from these regions and reducing the overall \cii emission from high-redshift galaxies. 

For a lower metallicity, and thus likely also lower dust content, UV photons have a longer mean free path. 
Also, it has been suggested that the binary stellar populations might yield an increased production of UV-photons (e.g. \citealt{ma16,stanway16,gotberg20}; however, see recent results from \citealt{ma20}). 
An increased intensity of the UV-radiation field would also impact the ionisation state of carbon, for example resulting in an increased fraction of double- or triple-ionised carbon (the ionisation potential for double ionisation is 24.4\,eV). Examples of double-ionised, and triple-ionised carbon have been found \citep[e.g.][]{Stark2015,smit17} for $z>5$ galaxies. 
We note that our continuum non-detection could rule out a dust-rich ISM scenario.  

Short-term variations in the star formation have also been suggested to impact the \cii luminosity relative to the estimated SFR \citep[e.g. ][]{Vallini2015,Ferrara2019}. For example, upward deviations from the Kennicutt-Schmidt star formation relation could be seen as a starburst-like phase, and that in turn can suppress the \cii emission \citep{Ferrara2019}. 

In addition, according to the CLOUDY simulations ran by \citet{Harikane2020}, the ionisation parameter may be the most important driver for a possible low \cii content at a given SFR.

For a further understanding of the non-detections from our stacking analysis, either deeper data for individual sources of the sample or a multiple-line analysis would be needed.  In terms of the latter, observations of the \oiii and \ciii lines specifically would help break the degeneracy   \citep[e.g.][]{Ferrara2019,Vallini2020}.

\section{Summary}

Through our analysis we performed a spectral stacking analysis of \cii\!, in 52 gravitationally lensed galaxies at $z\sim6$ using data from the ALCS.

We analysed both the full sample as well as three sub-samples, one containing only the sources with good spectroscopic redshifts, one with only sources with the higher SFR, and the last one with only sources present in both sub-samples. For each (sub-)sample we also performed a continuum stacking analysis. We performed both weighted-average and median stacking analyses and none yielded a detection. 
We derive $3\sigma$ upper limits on the \lcii\ of $(1-9)\times10^{7}$\,L$_\odot$. Compared to the local \lcii-SFR relation, this implies a deviation towards lower \lcii, similar to previous results for individual detections of $z>6$ lensed galaxies.  

The continuum stacking analyses also yielded non-detections and we derived upper limits on the far-infrared luminosity for two different modified blackbody temperatures of 35 and 45 K. The resulting upper limits vary between $0.4\times10^{10}$\,L$_\odot$ and $2.6\times10^{10}$\,L$_\odot$ depending on the temperature and sub-sample. From these we estimated upper limits on the SFR, which suggest that less than half of the star formation is dust obscured.  

We discuss potential source of bias in our analysis. A spectral stacking analysis relies on good knowledge of the redshift, and for example possible systematic offsets between \cii and \lya\ redshifts, could reduce the performance of the stacking analysis. Sample inhomogeneity could also lead to a biased result if not all objects follow the same relationship.  Systematic uncertainties on derived magnification and SFR of the sample sources can impact the interpretation of our results and the comparison to the \lcii-SFR relation. Finally, as suggested by modelling \citep[e.g.][]{Ferrara2019,Vallini2020}, several physical parameters could affect the \lcii.  For example, metallicity or gas density distribution could result in lower \cii excitation, and hence lower \lcii. 

Despite these potential biases, our analysis is the first large-scale analysis of \cii in faint lensed galaxies at the end of the EoR. Our stacking analysis allows us to reach very low RMS (of order of 1mJy/beam in 60\kms\,channels) and, being based on an analysis of 52 objects, our non-detection has a high statistical significance.

\begin{acknowledgements}

We thank the anonymous referee for their helpful comments. This paper makes use of the ALMA data:  ALMA \#2018.1.00035.L, \#2013.1.00999.S, and \#2015.1.01425.S. ALMA is a partnership of the ESO (representing its member states), NSF (USA) and NINS (Japan), together with NRC (Canada), MOST and ASIAA (Taiwan), and KASI (Republic of Korea), in cooperation with the Republic of Chile. The Joint ALMA Observatory is operated by the ESO, AUI/NRAO, and NAOJ. KK acknowledges support from the Swedish Research Council (2015-05580), and the Knut and Alice Wallenberg Foundation. K. Kohno acknowledges the JSPS KAKENHI Grant Number JP17H06130 and the NAOJ ALMA Scientific Research Grant Number 2017-06B. FEB acknowledges support from  ANID grants CATA-Basal AFB-170002,  FONDECYT Regular 1190818, 1200495 and Millennium Science Initiative ICN12\_009. DE acknowledges support from a Beatriz Galindo senior fellowship (BG20/00224) from the Ministry of Science and Innovation. GEM and FV acknowledge  the Villum Fonden research grant 13160 “Gas to stars, stars to dust: tracing star formation across cosmic time” and the Cosmic Dawn Center of Excellence funded by the Danish National Research Foundation under then grant No. 140. FV acknowledges support from the Carlsberg Foundation Research Grant CF18-0388 “Galaxies: Rise and Death”.
\end{acknowledgements}

\bibliography{myBib}

\begin{appendix} %First appendix
\section{Median stacks velocity integrated maps}
Because the results from the median stacks are qualitatively the same results as the mean stacking results --the main difference between mean and median in Figure \ref{CIISFR} being mostly due to the different magnifications and SFR-- we decided to put here the velocity integrated maps from our median stacking analyses.

\begin{figure}
   \centering
   \includegraphics[width=\hsize]{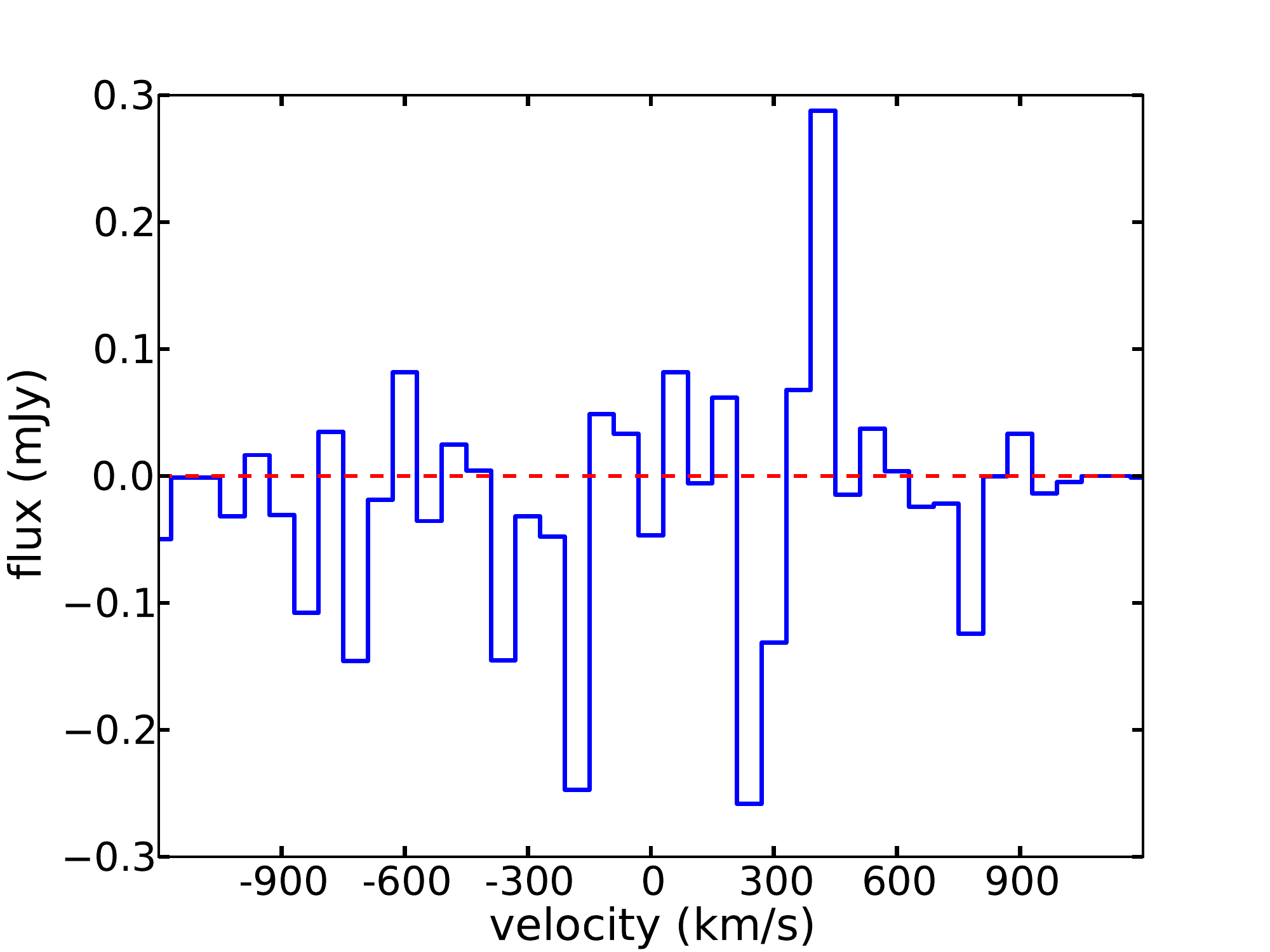}
      \caption{Spectrum extracted from a circular region of radius=1.12 (7 pixels) arcsec centred on the median stacked cube of the entire sample.}
         \label{CIIMedian}
   \end{figure}

\begin{figure*}
   \centering
    \includegraphics[width=0.9\columnwidth]{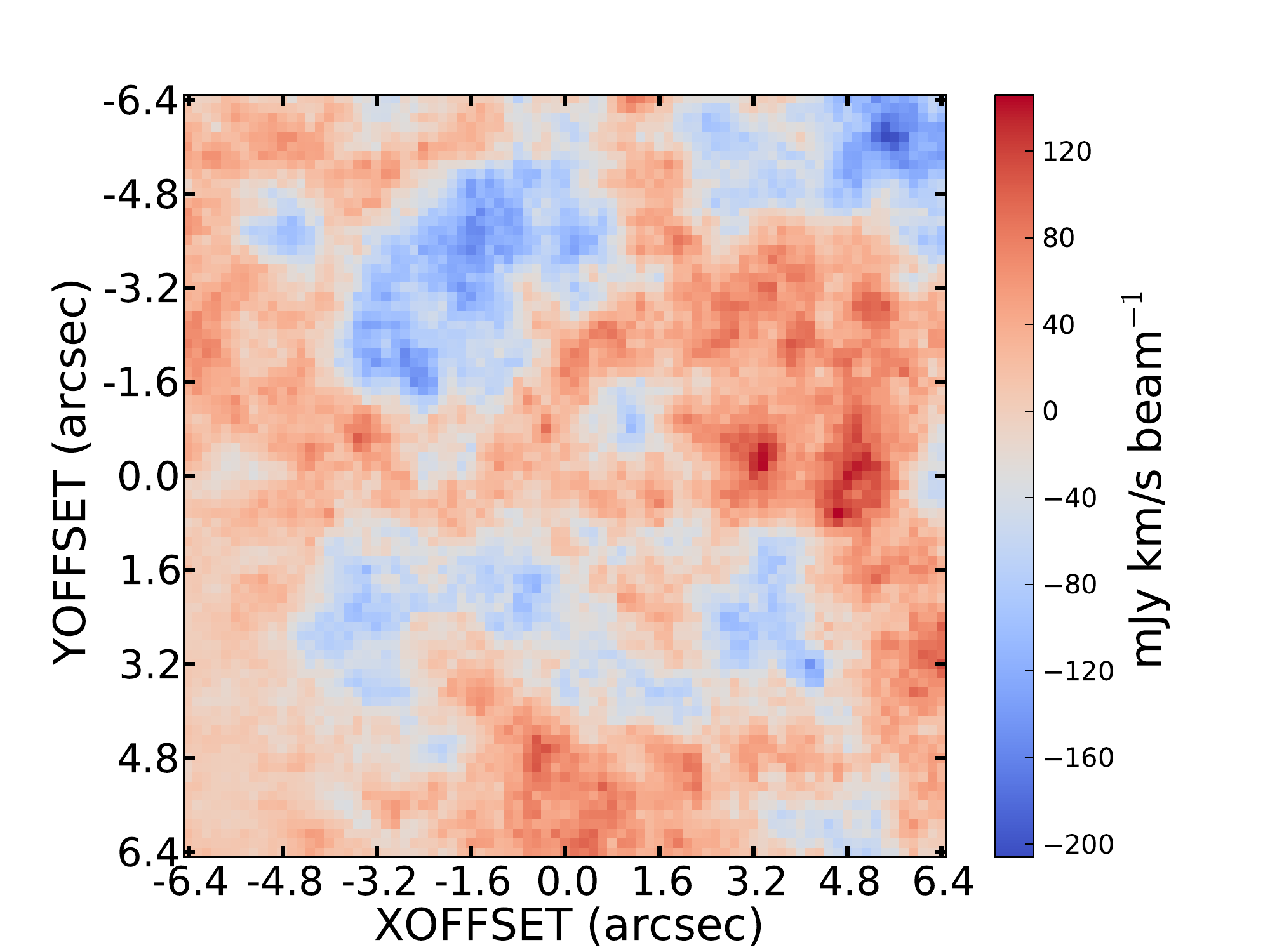}   
    \includegraphics[width=0.9\columnwidth]{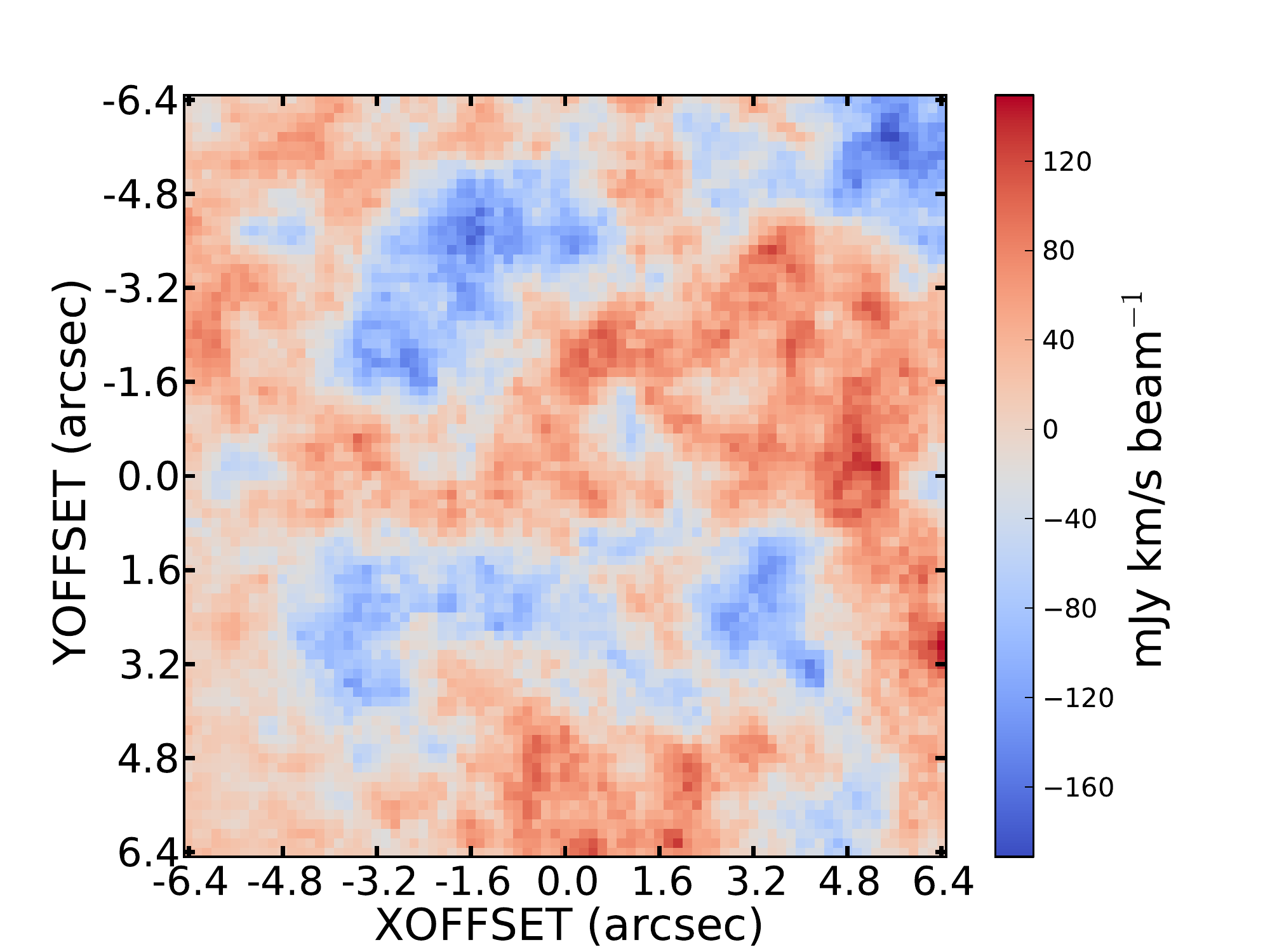}
    \includegraphics[width=0.9\columnwidth]{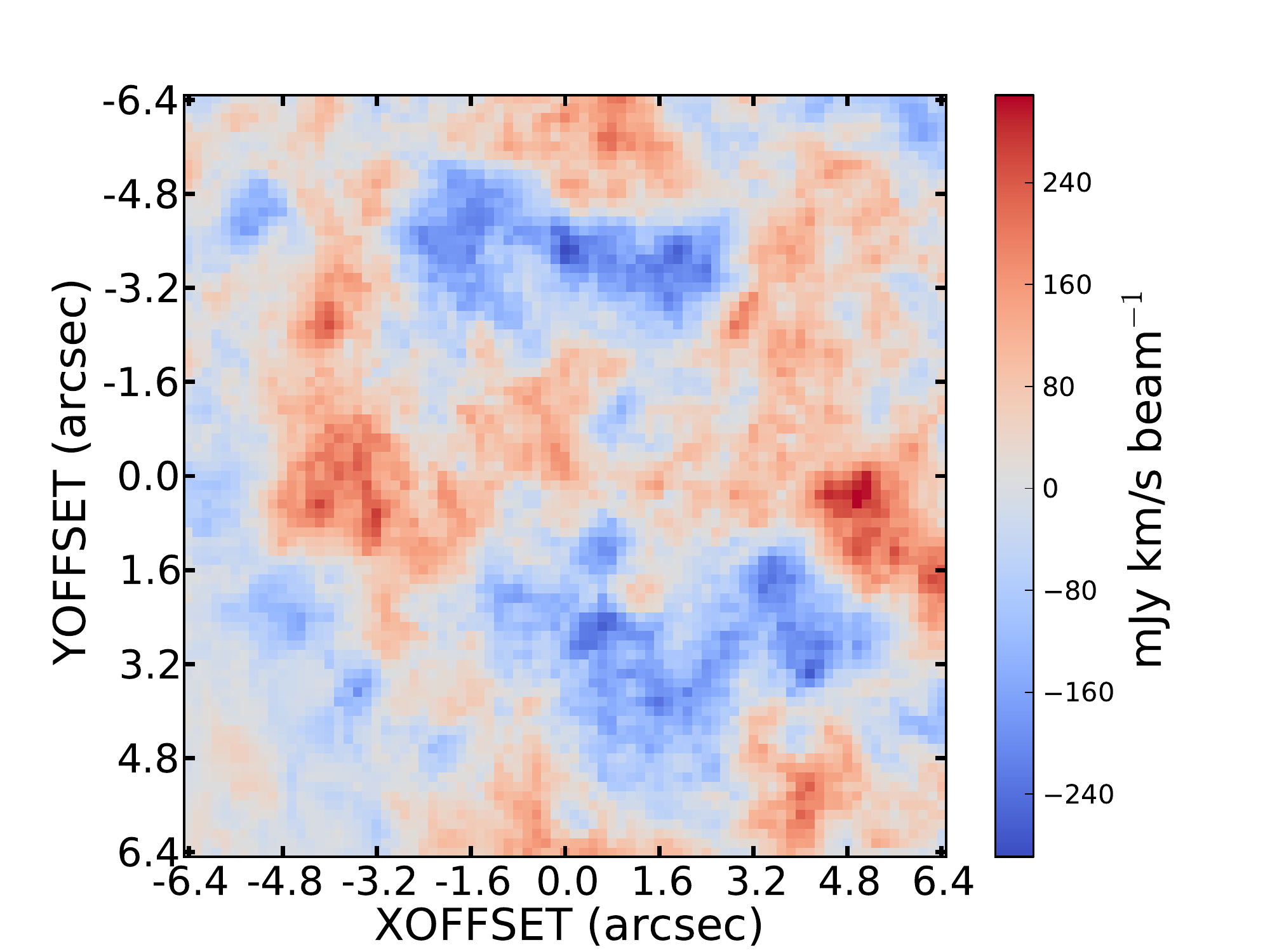}   
    \includegraphics[width=0.9\columnwidth]{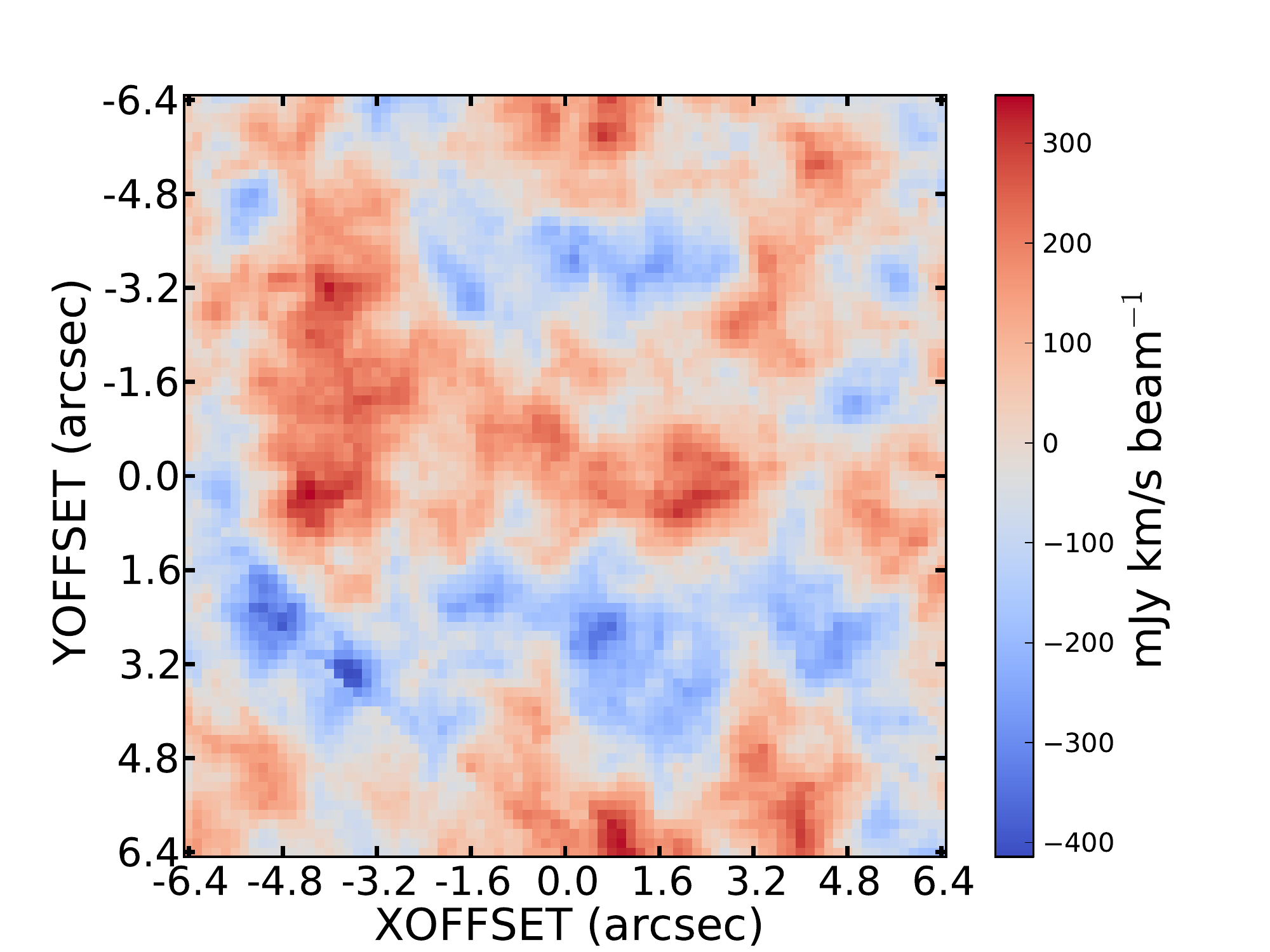}
      \caption{Velocity integrated flux maps of the median spectral stacks, obtained by collapsing a channel width of 540\,\kms\,centred on the stack centre. The artificial arcs featured on the sides of the figure are due to stacked sources located close to the edge of the cubes. From left to right and top to bottom, full-sample, good-$z$ sub-sample, high-SFR sub-sample and SFR+$z$ sub-sample.)
              }
         \label{CIIMedian_combined}
   \end{figure*}

\begin{figure*}
   \centering
    \includegraphics[width=0.9\columnwidth]{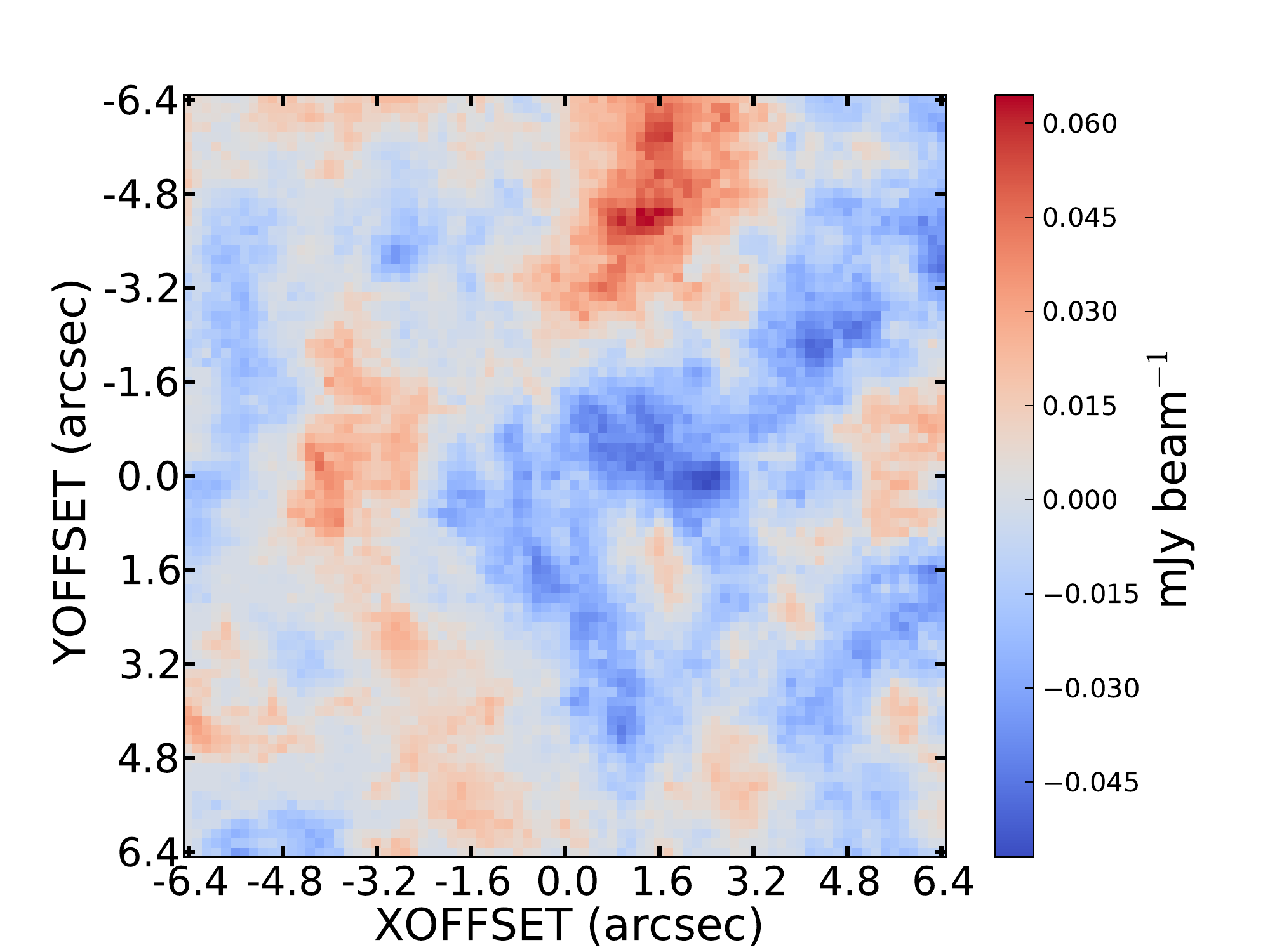}   
    \includegraphics[width=0.9\columnwidth]{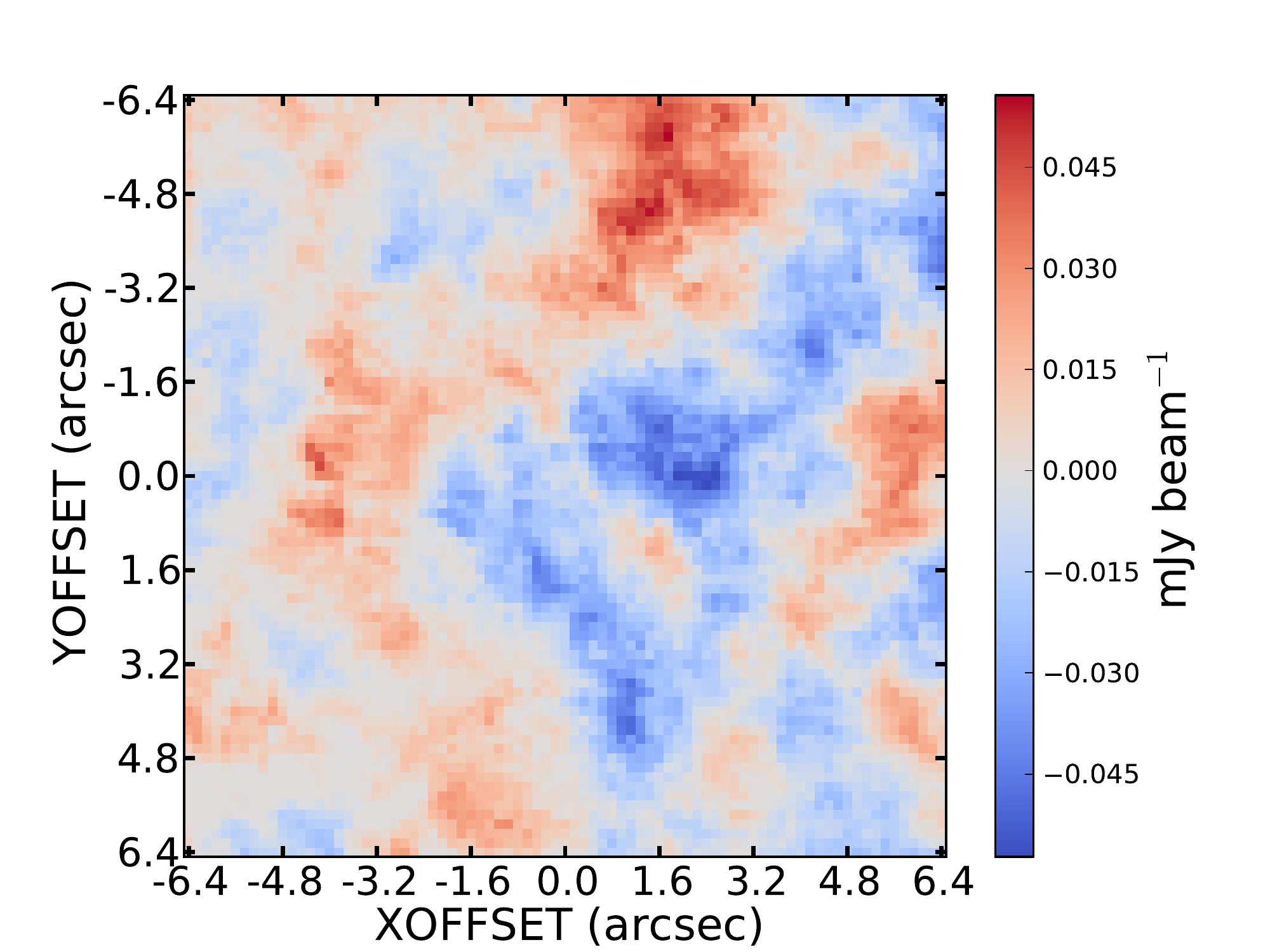}
    \includegraphics[width=0.9\columnwidth]{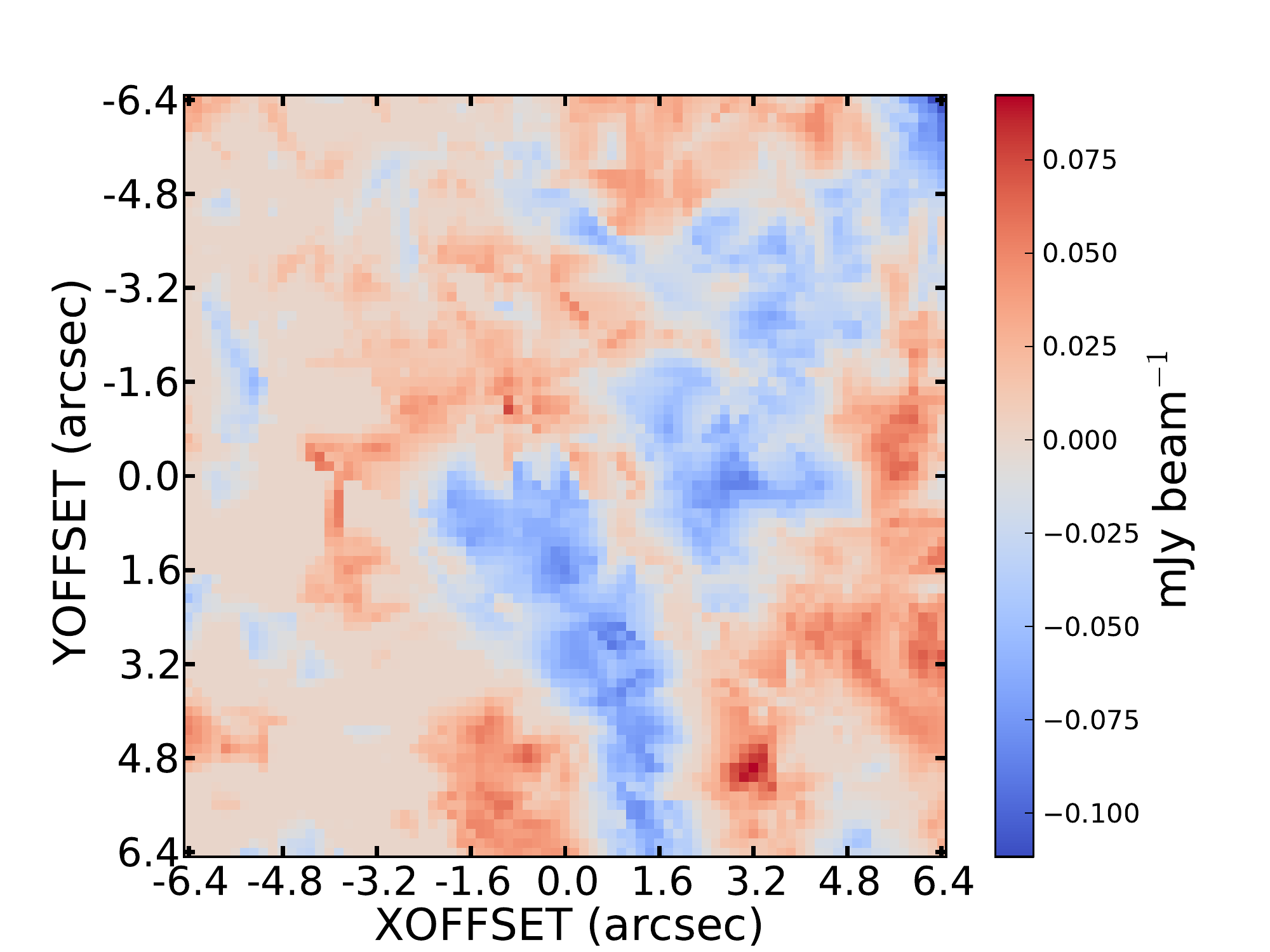}   
    \includegraphics[width=0.9\columnwidth]{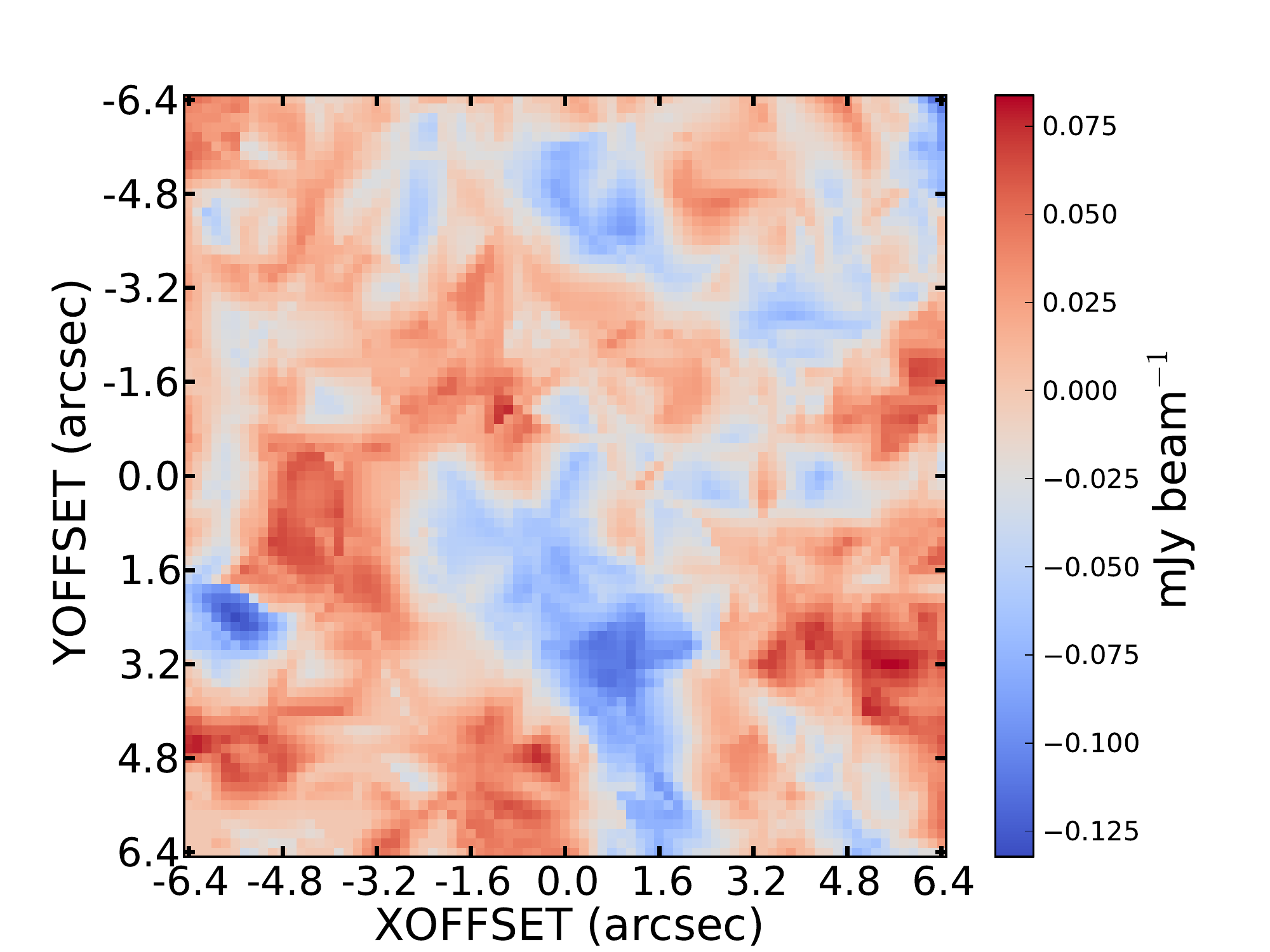}
      \caption{Continuum maps of the median continuum stacks. The artificial arcs featured on the sides of the figure are due to stacked sources located close to the edge of the cubes. From left to right and top to bottom, full-sample, good-$z$ sub-sample, high-SFR sub-sample and SFR+$z$ sub-sample. }
         \label{ContinuumMedian_combined}
   \end{figure*}

\begin{figure*}
   \centering
    \includegraphics[width=0.9\columnwidth]{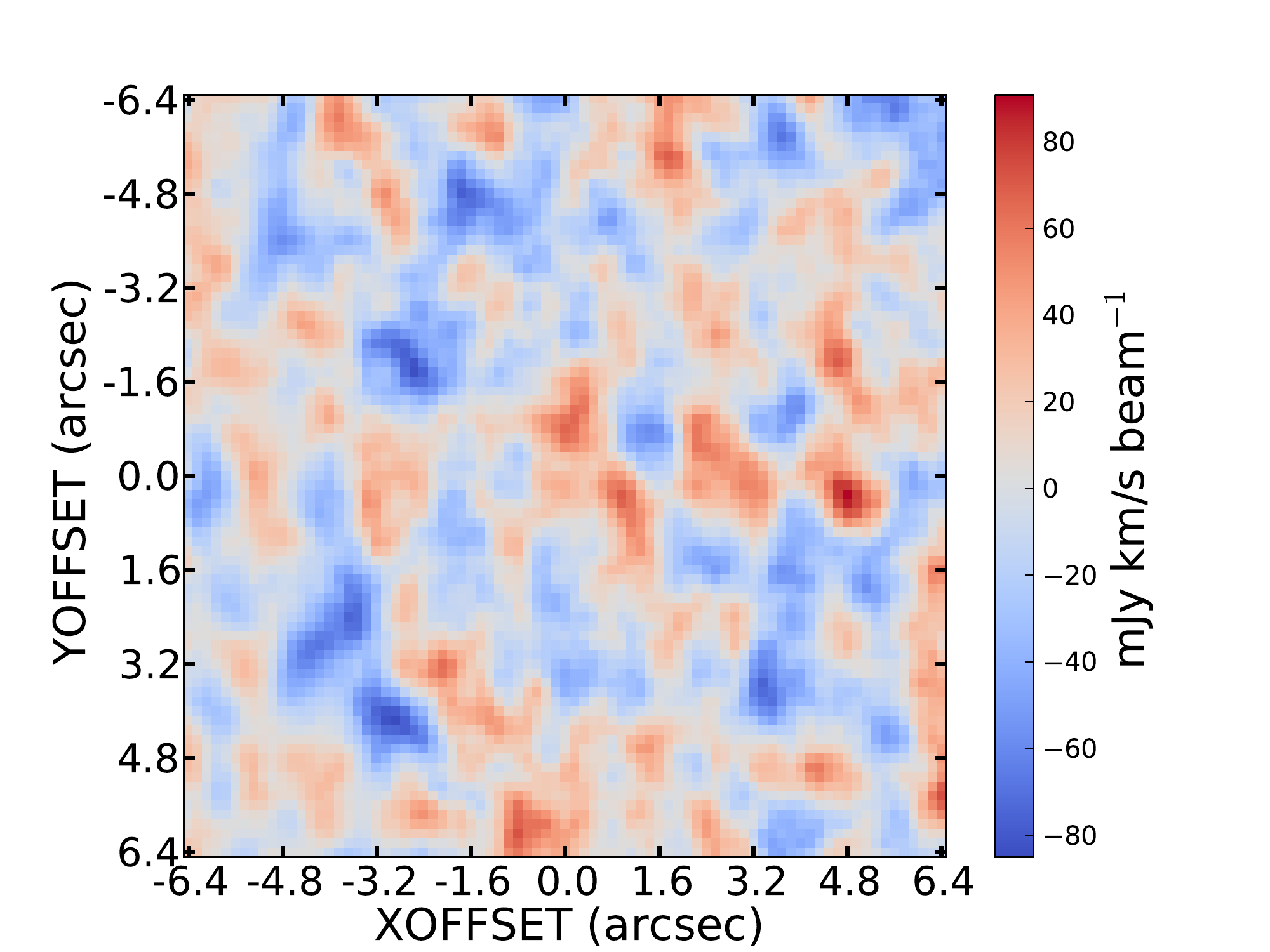}   
    \includegraphics[width=0.9\columnwidth]{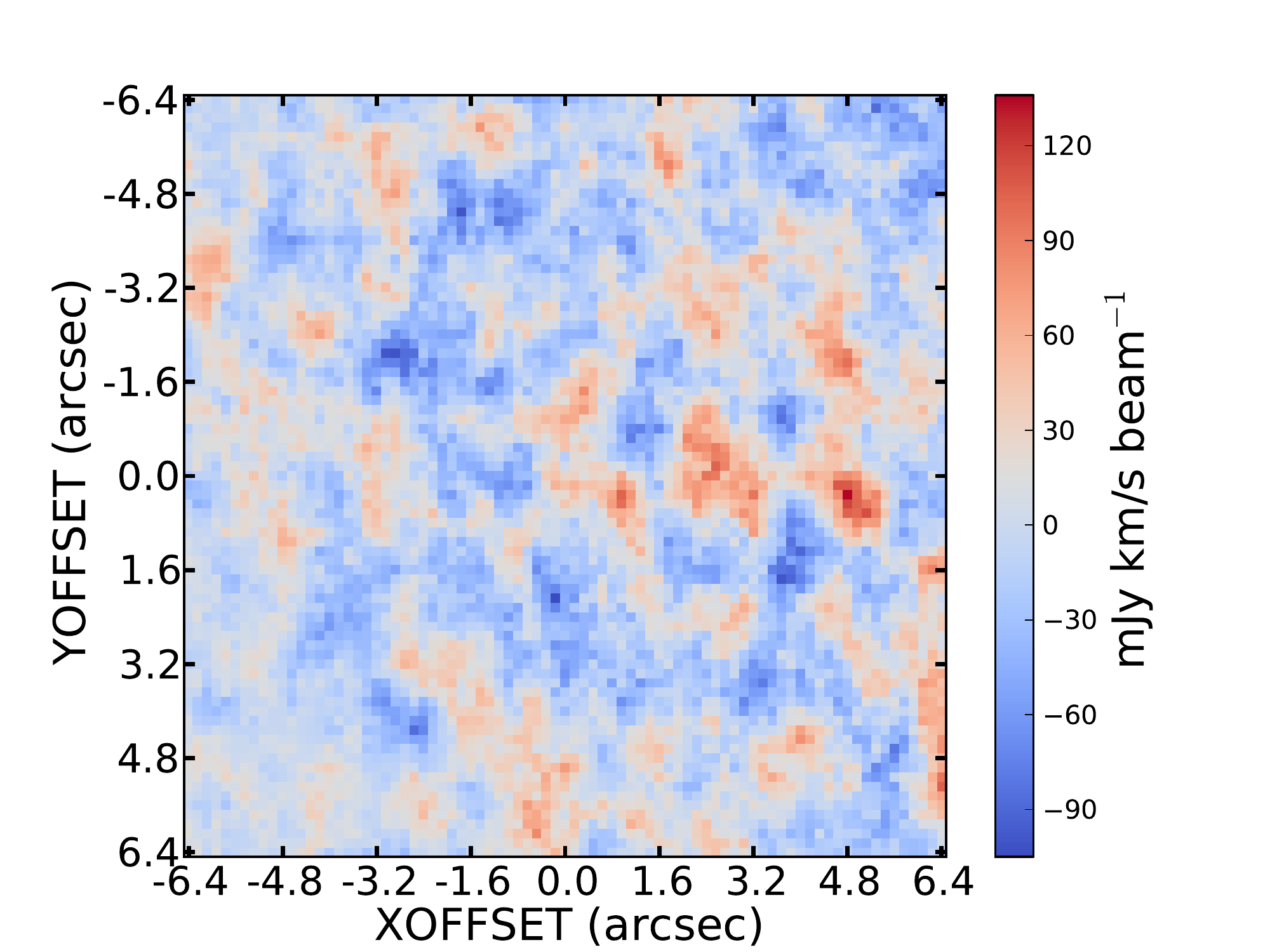}
      \caption{Velocity integrated flux maps of the mean (left) and median (right) spectral stacks of the cubes in their native resolution (not tapered), obtained by collapsing a channel width of 540\,\kms\,centred on the stack centre. The artificial arcs featured on the sides of the figure are due to stacked sources located close to the edge of the cubes. Both stacks are of the full-sample.
              }
         \label{CII_nativeRes}
   \end{figure*}

\section{Samples source descriptions and properties}

\longtab[1]{%\Small
\begin{longtable}{cccccccc}
\caption{Sample source description. } \\

%\hline
\hline

R.A.$^{(a)}$&Dec.$^{(b)}$& Redshift$^{(c)}$& Redshift Flag$^{(d)}$ & catalogue$^{(e)}$ & weight$^{(f)}$&Multiple$^{(g)}$&sub-sample$^{(h)}$ \\
 (deg) & (deg) & & /precision & & & & \\
\hline
\endfirsthead
\caption{continued.}\\
\hline\hline
\hline
\endhead
\hline
\endfoot
    \multicolumn{6}{l}{Abell\_2744}\\
    3.5938&-30.4154&6.588& 2 &  R & 0.445 &-& 3\\
    3.5801&-30.4079&6.556& 1 &  R & 0.102 &-&0\\
    3.5831&-30.4119&6.519& 1 &  R & 0.092 &-&0\\
    3.5769&-30.3863&6.457& 0.01715 &  AD & 0.482 & - &0\\
    3.5983&-30.4178&6.400& 0.01555 &  AD & 0.049 & - &2\\
    3.5728&-30.4145&6.477& 0.0185 &  AD & 0.08 & - & 2\\
      [0.3cm]
      \hline
    \multicolumn{6}{l}{MACS\_J0416.1-2403}\\
    64.0508&-24.0664&6.148& 3 &  R & 0.097 &96.3 , 97.3 & 1\\
    64.0482&-24.0624&6.147& 3 &  R & 0.122 &96.2&3\\
    64.0429&-24.0572&6.145& 3 &  R & 0.098 &-&1\\
    64.0435&-24.0590&6.147& 3 &  R & 0.106 &96.1, 97.1&1\\
    64.0478&-24.0621&6.147& 3 &  R & 0.121 &97.2 & 3\\
    64.0260&-24.0891&6.15& 3 &  R & 0.126 &-&1\\
    64.0318&-24.0889&6.405& 0.01555 &  AD & 0.76 & - &0\\
    64.0325&-24.0557&5.996& 0.0195 &  AD & 0.43 & - & 2\\
    64.0459&-24.0742&6.457& 0.01 &  AD & 0.867 & -&0\\
    64.0537&-24.0711&6.385& 0.01 &  AD & 0.833 & -&0\\
    64.0330&-24.0662&6.455& 0.0065 &  AD & 1.0 & - & 2\\
    64.0439&-24.0563&6.161& 0.011 &  AD & 0.079 & -&0\\
    64.0493&-24.0710&6.147& 2 &  R & 0.044 & 71.3&1\\
    64.0398&-24.0669&5.998& 2 &  R & 0.569 & 77.2&1\\
    64.0364&-24.0622&6.066& 2 &  R & 0.169 & 78.2&1\\
    64.0484&-24.0736&6.066& 2 &  R & 0.057 &78.3&1\\
    64.0458&-24.0741&5.998& 2 &  R & 0.496 &77.1&1\\
    64.0517&-24.0695&6.561& 2 &  R & 0.047 &-&1\\
    64.0556&-24.0638&6.074& 2 &  R & 0.06 &-&1\\
    64.0520&-24.0705&6.067& 2 &  R & 0.043 &-&1\\
    64.0460&-24.0582&6.498& 1 &  R & 0.161 &-&0\\
    64.0410&-24.0641&6.066& 3 &  R & 0.196 &78.1&1\\
    64.0433&-24.0629&6.147& 2 &  R & 0.12 &71.2&1\\
    64.0389&-24.0606&6.147& 1 &  R & 0.106 &71.1&0\\
      [0.3cm]
      \hline  
    \multicolumn{6}{l}{Abell\_S1063}\\
    342.1841&-44.5353&6.107& 1LIN &  NED & 0.463 & 14.4 &1\\
    342.1924&-44.5157&6.50& SPEC &  NED & 0.417 & - & 3\\
    342.1909&-44.5375&6.107& SPEC &  HFFDS & 0.202 & 14.1 & 3\\
    342.1811&-44.5346&6.107& SPEC &  HFFDS & 0.458 & 14.2 &1\\
    342.1841&-44.5317&6.107& SPEC &  HFFDS & 0.451 & 14.5 & 3\\
    342.1890&-44.5300&6.107& SPEC &  HFFDS & 0.455 & 14.3 & 3\\
      [0.3cm]
      \hline
    \multicolumn{6}{l}{MACS\_J1149.5+2223}\\
    177.3793&22.4033&6.478& 0.0185 &  AD & 0.081 & - & 2\\
      [0.3cm]
      \hline
    \multicolumn{6}{l}{MACS1206.2-0847}\\
    181.5511&-8.8037&6.014& 3 &  R & 0.591 & - &1\\
    181.5533&-8.7988&6.063& 2 &  R & 0.347 & 27.1&1\\
    181.5517&-8.7992&6.06& 2 &  R & 0.37 & 27.2&1\\
      [0.3cm]
      \hline
    \multicolumn{6}{l}{RXJ\_1347-1145}\\
    206.8921&-11.7483&5.980& 1 &  R & 0.324 & - & 2\\
    206.8762&-11.7386&6.457& 3 &  R & 0.344 & - & 3\\
    206.8832&-11.7563&6.566& 3 &  R & 0.341 & - &1\\
      [0.3cm]
      \hline  
    \multicolumn{6}{l}{MACS0329.7-0211}\\
    52.4258&-2.1898&6.011&  2 & R & 0.216 & - &1\\
    52.4312&-2.1914&6.026& 2 &  R & 0.103 & 9.1 &1\\
    52.4202&-2.1945&6.026& 2 &  R & 0.248 & 9.2 &1\\
    52.4169&-2.1977&6.1& 2 &  C & 0.259 & - &0\\
    52.4281&-2.1964&6.011& 2 &  R & 0.246 & - &1\\
    52.4310&-2.2044&6.075& 1 &  R & 0.06 &-&0\\
      [0.3cm]
      \hline
    \multicolumn{6}{l}{MACS1115.9+0129}\\
    168.9553&1.4993&6.057& 9 & C & 0.054 & - & 3\\
      [0.3cm]
      \hline
    \multicolumn{6}{l}{Abell 383}\\
    42.0136&-3.5263&6.027& Abell383 & R11+KK & 0.263 & 5.1 & 3\\
    42.0192&-3.5329&6.027& Abell383 & R11 & 0.218 & 5.2 & 3\\

\label{table:sample}
\end{longtable}
\tablefoot{
$(a)$ Right Ascension coordinate of the source (degrees); $(b)$ Declination coordinate of the source (degrees); $(c)$ Redshift of the source ;$(d)$ redshift confidence flag, sources from Richard: 3= highly certain, no doubts, 2= very good, usually one very bright emission line or multiple faint features but matching in redshift, 1= uncertain redshift. Caminha: (2) likely ; (3) secure measurement ; (9) single line measurement. NED: (SPEC)  secure spectroscopic redshift [$\sim100\%$ reliability], (1LIN)  single-line [$>90\%$ reliability]. ASTRODEEP: Photometric redshift precision ($\Delta z <0.02$) ; $(e)$ Source catalogue provider. R stands for \citet{Richard2020}, AD for ASTRODEEP, HFFDS for Hubble Frontier Field DeepSpace, C for \citet{Caminha2019}, R11 for \citet{Richard2011} and K for \citet{Knudsen2016}; $(f)$ Source weight (normalised, $1/\sigma^2$) ; $(g)$ Image multiplicity, multiple images of the same source are indicated as x.y where x is the source number and y is the image number (from \citet{Karman2017,Richard2020}) ; $(h)$ sub-sample: 0 if the source does not belong to any sub-sample, 1 if it belongs to the good-$z$ sub-sample, 2 if it belongs to the high-SFR sub-sample, 3 if it belongs to both the good-$z$ and high-SFR sub-sample.
}
}

\longtab[2]{
\begin{longtable}[c]{ccccc}
\caption{Physical properties of the sources.} \\
%\hline 
\hline
R.A.$^{(a)}$&Dec.$^{(b)}$& Redshift$^{(c)}$&Magnification$^{(d)}$&Star Formation Rate$^{(e)}$ \\
(deg) & (deg) &&&(M$_\odot$/year)\\
\hline
\endfirsthead
\caption{continued.}\\
\hline\hline
\hline
\endhead
\hline
\endfoot
\multicolumn{5}{l}{Abell\_2744}\\
3.5938&-30.4154&6.588& $4.06^{+0.06}_{-0.06}$ & $6.2^{+6.0}_{-2.4}$ \\
3.5801&-30.4079&6.556& 2.70 R& -\\
3.5831&-30.4119&6.519& 3.00 R& -\\
3.5769&-30.3863&6.457& $4.50^{+0.10}_{-0.10}$ & $1.4^{+1.3}_{-0.7}$\\
3.5983&-30.4178&6.4& $2.80^{+0.05}_{-0.06}$ & $2.2^{+2.4}_{-1.3}$\\
3.5728&-30.4145&6.477& $1.81^{+0.14}_{-0.13}$ & $5.3^{+9.6}_{-4.1}$\\
  [0.3cm]
  \hline
\multicolumn{5}{l}{MACS\_J0416.1-2403}\\
64.0508&-24.0664&6.148&  $8.47^{+0.61}_{-0.45}$ & 0.4 R\\
64.0482&-24.0624&6.147& $31.85^{+0.91}_{-1.11}$ & $2.1^{+5.0}_{-1.6}$ \\
64.0429&-24.0572&6.145& $4.04^{+0.17}_{-0.10}$ & $0.6^{+0.4}_{-0.3}$ \\
64.0435&-24.0590&6.147& 4.81 R& 0.6 R\\
64.0478&-24.0621&6.147& $23.82^{+0.55}_{-0.59}$ & $3.3^{+3.1}_{-1.5}$ \\
64.0260&-24.0891&6.15& 9.63 R& -\\
64.0318&-24.0889&6.405& $2.83^{+0.06}_{-0.08}$ & $1.3^{+1.8}_{-0.9}$\\
64.0325&-24.0557&5.996& $1.79^{+0.02}_{-0.03}$ & $4.9^{+4.2}_{-2.8}$\\
64.0459&-24.0742&6.457& $3.27^{+0.04}_{-0.05}$ & $1.4^{+1.7}_{-0.8}$ \\
64.0537&-24.0711&6.385&  1.81 AD & 0.7 AD\\
64.0330&-24.0662&6.455& $4.49^{+0.18}_{-0.15}$ & $2.4^{+4.5}_{-1.6}$ \\
64.0439&-24.0563&6.161& $3.82^{+0.16}_{-0.08}$ & $2.0^{+4.3}_{-1.5}$\\
64.0493&-24.0710&6.147& 2.23 R& -\\
64.0398&-24.0669&5.998& 2.21 R& - \\
64.0364&-24.0622&6.066& 3.37 R & $1.1^{+0.4}_{-0.3}$ \\
64.0484&-24.0736&6.066& 2.08 R& -\\
64.0458&-24.0741&5.998& 3.20 R& - \\
64.0517&-24.0695&6.561& 2.01 R& -\\
64.0556&-24.0638&6.074& 2.12 R& -\\
64.0520&-24.0705&6.067& 1.85 R & $0.5^{+0.7}_{-0.2}$ \\
64.0460&-24.0582&6.498& 8.34 R & $1.3^{+2.2}_{-0.6}$ \\
64.0410&-24.0641&6.066& 2.82 R & $0.9^{+1.8}_{-0.5}$ \\
64.0433&-24.0629&6.147& 3.69 R & 1.3 R\\
64.0389&-24.0606&6.147& 4.05 R & $1.5^{+0.8}_{-0.4}$ \\
  [0.3cm]
  \hline  
\multicolumn{5}{l}{Abell\_S1063}\\
342.1841&-44.5353&6.107& 16.73  & -\\
342.1924&-44.5157&6.5& $3.96^{+0.10}_{-0.08}$ & $5.4^{+0.8}_{-0.7}$\\
342.1909&-44.5375&6.107& $4.70^{+0.23}_{-0.18}$ & $12.3^{+2.0}_{-1.2}$\\
342.1811&-44.5346&6.107& $4.41^{+0.24}_{-0.19}$ & $1.8^{+212.9}_{-1.8}$\\
342.1841&-44.5317&6.107& 1.77 HFFDS & 125.9 HFFDS\\
342.1890&-44.5300&6.107& 5.35 HFFDS & 23.4 HFFDS\\
  [0.3cm]
  \hline
\multicolumn{5}{l}{MACS\_J1149.5+2223}\\
177.3793&22.4033&6.478& $2.55^{+0.13}_{-0.10}$ & $4.7^{+2.4}_{-1.9}$\\
  [0.3cm]
  \hline
\multicolumn{5}{l}{MACS1206.2-0847}\\
181.5511&-8.8037&6.014& 72.90 R& -\\
181.5533&-8.7988&6.063& $13.00^{+12.82}_{-13.20}$ & $0.2^{+1.0}_{-0.2}$\\
181.5517&-8.7992&6.06& 6.88 R&- \\
  [0.3cm]
  \hline
\multicolumn{5}{l}{RXJ\_1347-1145}\\
206.8921&-11.7483&5.98& $2.89^{+2.85}_{-2.94}$  & $6.5^{+3.4}_{-3.2}$\\
206.8762&-11.7386&6.457& $2.19^{+2.23}_{-2.33}$ & $10.5^{+2.7}_{-1.1}$\\
206.8832&-11.7563&6.566& 12.29 R& - \\
  [0.3cm]
  \hline  
\multicolumn{5}{l}{MACS0329.7-0211}\\
52.4258&-2.1898&6.011& 11.52 R& -\\
52.4312&-2.1914&6.026& 3.51 R& -\\
52.4202&-2.1945&6.026& 4.93 R& -\\
52.4169&-2.1977&6.1& $15.51^{+13.82}_{-15.55}$ & $0.4^{+0.6}_{-0.2}$\\
52.4281&-2.1964&6.011& 10.71 R& -\\
52.4310&-2.2044&6.075& 3.72 R& -\\
  [0.3cm]
  \hline
\multicolumn{5}{l}{MACS1115.9+0129}\\
168.9553&1.4993&6.057& $2.86^{+2.74}_{-2.89}$ & $76.5^{+43.6}_{-31.8}$\\
  [0.3cm]
  \hline
\multicolumn{5}{l}{Abell 383}\\
42.0136&-3.5263&6.027& $5.15^{+4.59}_{-5.09}$ & $24.0^{+15.1}_{-11.2}$\\
42.0192&-3.5329&6.027& $3.27^{+3.13}_{-3.40}$ & $8.4^{+13.3}_{-5.1}$\\

\label{table:physics}
\end{longtable}
\tablefoot{
IDs are placed next to values when they are extracted from the literature, R stands for \citet{Richard2020}, AD for ASTRODEEP and HFFDS for Huble Frontier Field DeepSpace. Values with no mention have been computed through our own analysis presented in subsection \ref{sampleProperties} $(a)$ Right Ascension coordinate of the source (degrees); $(b)$ Declination coordinate of the source (degrees); $(c)$ Redshift of the source ;$(d)$ Source lensing magnification ; $(e)$ Source SFR. When not derived from the literature, SFR were derived using the UV luminosity and corrected for dust attenuation (see Sect. \ref{sampleProperties}).}
}

\end{appendix}
% WARNING
%-------------------------------------------------------------------
% Please note that we have included the references to the file aa.dem in
% order to compile it, but we ask you to:
%
% - use BibTeX with the regular commands:
%   \bibliographystyle{aa} % style aa.bst
%   \bibliography{Yourfile} % your references Yourfile.bib
%
% - join the .bib files when you upload your source files
%-------------------------------------------------------------------

\end{document}